# Surface effects in nucleation

Nikolay V. Alekseechkin

Akhiezer Institute for Theoretical Physics, National Science Centre "Kharkov Institute of Physics and Technology", Akademicheskaya Street 1, Kharkov 61108, Ukraine

The classical nucleation theory (CNT) concept of a nucleus as a fragment of the bulk new phase fails for nanosized nuclei. An extension of CNT taking into account the properties of the transition region between coexisting bulk phases is proposed. For this purpose, the finite-thickness layer method which is an alternative to Gibbs' one is used; the transition region is considered as a separate (surface) phase. An equation for the nucleation work is derived which is basic for the multivariable theory of nucleation. Equations of equilibrium following from it are employed for considering the dependences of surface tension on radius and temperature for droplets; Kelvin's formula for the equilibrium vapor pressure is extended to small radii. It is shown that the ratio of the isothermal nucleation rate to that of CNT can achieve several orders of magnitude due to the curvature effect (the dependence of surface tension on curvature). The analysis of different dependences of the Tolman length on radius, $\delta(R)$, suggests that (i) the curvature effect is determined by the value of $\delta(0)$ which is positive and relates to the limiting (spinodal) vapor supersaturation and (ii) the function $\delta(R)$ decreases with increasing radius; at the same time, this effect is weakly sensitive to the form of the function $\delta(R)$ and its asymptotic value $\delta_\infty$. The second differential of the work is obtained as a quadratic form with contributions from both the bulk and surface phases. It is used for calculating the fluctuations of surface layer parameters such as the surface tension and the specific surface area as well as the fluctuations of nucleus parameters. The matrix of the mentioned quadratic form for a noncritical nucleus is found to differ essentially from the corresponding CNT matrix; it involves the off-diagonal elements related to the dependence of surface tension on temperature. As a consequence, the defined equilibrium temperature of a noncritical droplet differs from the vapor temperature and the calculated mean steady state overheat of droplets consists of the kinetic (due to the release of the condensation heat) and thermodynamic parts.

# 1. Introduction

CNT [1-3] is based on the so-called drop model: the work of formation of a new phase nucleus is represented as the sum of volume and surface parts. The surface part is proportional to the nucleus surface area; the proportionality factor is the surface tension of the planar interface. Thermodynamic parameters of the volume part are taken the same as for the bulk phase. In other words, the nucleus is represented as a small fragment of the bulk phase and the presence of the transition region (or the surface layer) between the new and old phases is neglected. At the same time, the typical radius of a critical nucleus is about 1 nm, so that the homogeneous (bulk) phase inside this nucleus is absent and it entirely consists of the inhomogeneous surface layer with thermodynamic parameters different from those of the bulk phase.

The development of the density functional theory [4-10] was intended to take into account the inhomogeneity of a new phase nucleus and thereby to improve the predictive ability of the nucleation theory; this new theory was named non-classical [4]. It made significant progress in the description of nucleus properties; density profiles, nucleation barriers, etc. were calculated. The density functional theory uses certain models for the structure of substance, in particular, a certain form of the intermolecular potential and therefore it is related to the first-principles theories [6].

At the same time, it is of interest to study the given problem within a thermodynamic approach for the following reasons. First, as is known, thermodynamic relations have a great generality. Even relations in the form of inequalities are useful for the analysis of processes. Second, the use of macroscopic kinetic equations for the description of nucleus evolution [11-16] in the multivariable nucleation theory [15-22] requires a thermodynamic expression for the nucleation work.

There are two thermodynamic approaches to studying surface phenomena: Gibbs' theory of capillarity [23] and the finite-thickness layer (FTL) method [24]. According to Gibb's approach, the real surface layer is replaced by the geometrical surface possessing the tension $\sigma$ and to which the excess energy, $E_\Sigma$, entropy, $S_\Sigma$, and numbers of particles (or masses), $N_\Sigma^i$, are related. In Ref. [11], the thermodynamics of the new-phase nucleus formation was considered within Gibb's approach. The excess entropy $S_\Sigma$ and number of particles $N_\Sigma$ of a single component system "nucleus and ambient phase" were represented as the sum of contributions from new phase $\alpha$ and old phase $\beta$ (the "one-sided" superficial quantities), $S_\Sigma = S_\Sigma^\alpha + S_\Sigma^\beta$, $N_\Sigma = N_\Sigma^\alpha + N_\Sigma^\beta$. Such separation is uniquely determined by setting the position of the geometrical dividing surface; it is of no importance in original Gibbs' work, since the mentioned system is considered in the equilibrium state, i.e. at equal temperatures of phases $\alpha$ and $\beta$, $T^\alpha = T^\beta$, and equal chemical potentials, $\mu^\alpha = \mu^\beta$. In Ref. [11], a noncritical nucleus is considered which is not in equilibrium with the ambient phase, accordingly, the temperatures and chemical potentials of phases $\alpha$ and $\beta$ are different. The above separation makes it possible to take into account each part, $S_\Sigma^\alpha$ and $S_\Sigma^\beta$ ($N_\Sigma^\alpha$ and $N_\Sigma^\beta$), with its own temperature (chemical potential) in the fundamental equations for energy. As a result, the following equation for the nucleation work was derived therein:

$$W = (\mu^\alpha - \mu^\beta)(N^\alpha + N_\Sigma^\alpha) + (T^\alpha - T^\beta)(S^\alpha + S_\Sigma^\alpha) - (P^\alpha - P^\beta)V + \sigma A \tag{1}$$

where $N^\alpha$ and $S^\alpha$ are the number of particles and entropy of bulk phase $\alpha$ in the nucleus; thus, $(N^\alpha + N_\Sigma^\alpha)$ and $(S^\alpha + S_\Sigma^\alpha)$ are the *true* values of the number of particles and entropy of the nucleus. $P^\alpha$ and $P^\beta$ are the pressures in each phase, $V$ and $A$ are the volume and surface area of the nucleus, respectively; the dividing surface (the surface of tension) is the nucleus boundary.

The adsorption equation at a fixed state of phase $\beta$ was obtained as [11]

$$(Ad\sigma)_{T^\beta,\mu^\beta} = -S_\Sigma^\alpha dT^\alpha - N_\Sigma^\alpha d\mu^\alpha \tag{2}$$

The condition of equilibrium, or the saddle-point condition, $(dW)_{T^\beta,P^\beta} = 0$ with account for Eq. (2) and the Gibbs-Duhem equation for phase $\alpha$ results in conventional equations of equilibrium: $T^\alpha = T^\beta$, $\mu^\alpha = \mu^\beta$, and $P^\alpha - P^\beta = 2\sigma / R_* \equiv P_L^*$, where $R$ is the nucleus radius; hereafter, asterisk denotes the saddle-point value of the corresponding quantity. So, near the saddle point, the work is

$$W = W_* + \frac{1}{2}(d^2W)_{T^\beta,P^\beta}^* = W_* + \frac{1}{2}\sum_{i,k} h_{ik}(\xi_i - \xi_i^*)(\xi_k - \xi_k^*) \tag{3}$$

where $\{\xi_i\}$ are the nucleus parameters.

The following equation is obtained [11] for the second differential of $W$ from Eq. (1):

$$(d^2W)_* = -\frac{P_L^*}{3V_*}(dV)^2 + N_*^\alpha \left[ dT^\alpha ds^\alpha - dP^\alpha d\upsilon^\alpha \right]_* + (d^2W)_*^\sigma \tag{4a}$$

where

$$(d^2W)^\sigma \equiv d\mu^\alpha dN_\Sigma^\alpha + dT^\alpha dS_\Sigma^\alpha + d\sigma dA \tag{4b}$$

and $s^\alpha = S^\alpha / N^\alpha$, $\upsilon^\alpha = V / N^\alpha$.

The approximation $(d^2W)^\sigma = 0$ together with Eq. (2) leads to equations

$$\frac{dS_\Sigma^\alpha}{dA} = \frac{S_\Sigma^\alpha}{A} \quad \text{and} \quad \frac{dN_\Sigma^\alpha}{dA} = \frac{N_\Sigma^\alpha}{A} \tag{5a}$$

having the solutions

$$S_\Sigma^\alpha = c_s A \text{ and } N_\Sigma^\alpha = c_N A \tag{5b}$$

with *constant* $c_s$ and $c_N$.

Eq. (4a) is basic for the multivariable nucleation theory; it was employed (without the surface term $(d^2W)^\sigma$) in Refs. [11-14] for studying the nucleation of bubbles and droplets in the multivariable approach. The negative term in Eq. (4a) corresponds to the unstable variable – the nucleus volume, whereas the term in square brackets is a positive definite quadratic form corresponding to stable variables (temperature $T^\alpha$ and molecular volume $\upsilon^\alpha$ or density $\rho^\alpha = 1/\upsilon^\alpha$). Thus, the latter describes the fluctuations of these quantities [11, 13]. The kinetic equations, or the equations of motion of a nucleus in the space of its variables [11-14], describe the coupling between the changes of these quantities and the volume change. The determination of the matrix **H** in Eq. (3) with the use of the explicit form of the second differential is needed for writing the kinetic equations for stable variables and thereby for calculating the equilibrium and steady-state distribution functions of nuclei and the nucleation rate.

It is seen from the foregoing that the neglect of the surface term $(d^2W)^\sigma$ in Eq. (4a) is equivalent to the condition of independence of superficial densities $c_s$ and $c_N$ of $A$, or, what is the same, of $R$. Apparently, this approximation is good only for a nucleus of a sufficiently large radius, whereas the keeping of the surface term is necessary for small nuclei. So, the multivariable theory without the term $(d^2W)^\sigma$ is simply an extension of the one-dimensional CNT to the case of greater number of variables; it has the same shortcoming of the latter – the parameters of bulk phase $\alpha$ are employed as nucleus parameters.

The aim of the present report is to take into account the effect of the surface term similar to Eq. (4b) on the thermodynamics and kinetics of nucleation. However, the mentioned above second approach – the FTL method – is employed here for this purpose. Eqs. (1)-(4) are presented here for comparison to the corresponding equations derived below within the given method. While Gibbs' approach operates with superficial quantities, the FTL method deals with real quantities having simple physical meaning and relating to the surface layer which is considered as a separate phase. As a consequence, the interpretation of results is greatly simplified in comparison with Gibbs' approach (especially in the case of a multicomponent system). The values of surface layer parameters included in thermodynamic equations can be estimated within the statistical mechanical approach, the density functional theory, and from computer simulations. The FTL method was first introduced by Dutch physicists [25-27], in particular, by Guggenheim [28] for planar interfaces. As applied to curved interfaces, this method was developed in detail by Rusanov [24].

The outline of the paper is as follows. In Section 2, the model is formulated and fundamental equations are written; the basic equation for the nucleation work is also derived here. The conditions of equilibrium obtained from this equation are applied to considering the dependences of surface tension on radius and temperature in Section 3. The applications of the second differential of the work are considered in Section 4; the fluctuations of surface layer parameters as well as the thermodynamics of a noncritical droplet are studied here. The application of thermodynamic results to the kinetics of droplet nucleation is considered in Section 5. The summary of results in Section 6 finalizes the paper.

## 2. Model and fundamental equations

## 2.1. Three-phase system

We consider the three-phase system consisting of new phase $\alpha$, parent phase $\beta$ and the surface layer between them which is phase $\sigma$ (Fig. 1). The surface of tension [24, 29, 30] is employed as a dividing surface. The following equation for its radius $R$ is obtained in the thermodynamic theory of a curved surface layer [24, 29]:

$$\frac{1}{2}\left(P^\alpha - P^\beta\right)R^3 = \int_{R^\alpha}^{R}(P^\alpha - P_t)r^2 dr + \int_{R}^{R^\beta}(P^\beta - P_t)r^2 dr = 0 \qquad (6)$$

where $P_t$ is the tangential component of the pressure tensor; $R^\alpha$ and $R^\beta$ are the surface layer boundaries (Fig. 1). The complex consisting of phases $\alpha$ and $\sigma$ is the *density fluctuation* [31, 32] (DF) within phase $\beta$ which was named "globule" by Gibbs [23].

When the surface of tension is used as a dividing surface, the work of deformation of a curved surface layer is represented as the sum of volume and surface contributions [24],

$$\delta w = -P^\alpha \delta V^{\alpha\sigma} - P^\beta \delta V^{\beta\sigma} + \sigma \delta A \qquad (7)$$

The last term in this equation is the work of stretching of some surface with tension $\sigma$. In other words, the mechanical similarity of the surface layer and a two-dimensional stretched film takes place only when the surface of tension is employed. With this choice of the dividing surface, Eq. (7) as well as fundamental thermodynamic equations do not involve the additional term related to the arbitrariness of the dividing surface position [24].

The natural question of the effective thickness $\tau = \tau^\alpha + \tau^\beta$ of the surface layer arises, where the components $\tau^\alpha = R - R^\alpha$ and $\tau^\beta = R^\beta - R$ are measured from the dividing surface. The introduction of this concept is possible due to the smallness of the radius of action of intermolecular forces, which causes rather rapid decline of the influence of one of the phases on any property of the neighboring phase. It should be noted that for different properties $\xi_i$, the thicknesses $\tau(\xi_i)$ are different; a formula for the dependence of any quantity $\xi_i$ on distance $l$ from some surface can be derived in statistical mechanics. In Ref. [24], such formulae are given for the profiles of density, $\xi_1 = \rho(l)$, and the tangential component of the pressure tensor, $\xi_2 = P_t(l)$, for the planar vapor-liquid interface; it is shown that the former function is much more steep than the latter. In a multicomponent system, the composition profiles are also considered.

Outside the effective thickness $\tau(\xi_i)$, the deviation of the local value of the quantity $\xi_i$ from its bulk value becomes negligible. It is natural to use fluctuations as a quantitative criterion: the above deviation has not to exceed the rms fluctuation of $\xi_i$:

$$\left|\xi_i(\tau^{\alpha(\beta)}) - \left\langle\xi_i^{\alpha(\beta)}\right\rangle\right| \leq \sqrt{\left\langle\left(\Delta\xi_i^{\alpha(\beta)}\right)^2\right\rangle} \qquad (8)$$

where $\left\langle \xi_i^{\alpha(\beta)} \right\rangle$ is the bulk value of $\xi_i$. Calculating $\tau^{\alpha(\beta)}(\xi_i)$ for different properties $\xi_i$ and then taking the maximum values of $\tau^{\alpha}$ and $\tau^{\beta}$, we find that the thickness $\tau = \tau^{\alpha} + \tau^{\beta}$ of the layer covers the region of essential change of all properties $\xi_i$; outside this layer, all properties coincide with the properties of bulk phases. The thickness $\tau$ is a state parameter and can be uniquely connected with temperature, pressure and other thermodynamic parameters. So, with this definition of the quantity $\tau$, the surface layer becomes an *autonomous* phase – the "surface phase".

Considering the system consisting of phases $\alpha$, $\beta$, and $\sigma$, we start from its total equilibrium and then vary the states of constituting phases. Three types of variations are of interest. (i) The states of all phases change, but they remain in equilibrium with each other, i.e. the total equilibrium of the system is retained. This situation corresponds to a critical nucleus in the ambient phase (the saddle point of the nucleation work). (ii) The state of one phase, $\sigma$, changes, which corresponds to the fluctuations of the surface layer. (iii) The states of two phases, $\alpha$ and $\sigma$, change, but they remain in equilibrium with each other (not with phase $\beta$). This situation corresponds to a noncritical nucleus; deviations from the total equilibrium are assumed to be small. Later, all these types are examined.

Considering the appearance of the DF, we mean that it occurs in a large amount of phase $\beta$ playing therefore the role of a thermostat, so that its parameters do not change at this event: temperature $T^{\beta}$, pressure $P^{\beta}$, and composition $\mathbf{x}^{\beta}$ are *constant*; hence, $\mu_i^{\beta}(T^{\beta}, P^{\beta}, \mathbf{x}^{\beta}) = const$, where $\mathbf{x}^{\beta} \equiv (x_1^{\beta}, x_2^{\beta}, ...., x_{n-1}^{\beta})$. The particles that make up the DF form the subsystem in phase $\beta$ which goes from one phase state (state 1) to another (the DF, state 2). As is known from thermodynamics [33, 34], the minimum (reversible) work done by the medium (phase $\beta$) in such a process is given by the following expression:

$$W = \Delta E - T^{\beta}\Delta S + P^{\beta}\Delta V \tag{9}$$

where the changes in energy, $\Delta E$, entropy, $\Delta S$, and volume, $\Delta V$, relate to the substance in the DF (Fig. 1). Since the state of phase $\beta$ does not change under the DF formation, these changes can be attributed to the *whole* system also. Eq. (9) is a consequence of the first and second laws of thermodynamics only.

## 2.2. Fundamental equations

The fundamental equations for energy are as follows. For a homogeneous (bulk) phase,

$$dE^{\alpha} = T^{\alpha}dS^{\alpha} - P^{\alpha}dV^{\alpha} + \sum_{i=1}^{n}\mu_i^{\alpha}dN_i^{\alpha} \tag{10a}$$

$$E^{\alpha} = T^{\alpha}S^{\alpha} - P^{\alpha}V^{\alpha} + \sum_{i=1}^{n}\mu_i^{\alpha}N_i^{\alpha} \tag{10b}$$

and the same equations for phase $\beta$.

For a curved surface layer, in view of Eq. (7),

$$dE^\sigma = T^\sigma dS^\sigma - P^\alpha dV^{\alpha\sigma} - P^\beta dV^{\beta\sigma} + \sigma dA + \sum_{i=1}^{n} \mu_i^\sigma dN_i^\sigma \tag{11a}$$

$$E^\sigma = T^\sigma S^\sigma - P^\alpha V^{\alpha\sigma} - P^\beta V^{\beta\sigma} + \sigma A + \sum_{i=1}^{n} \mu_i^\sigma N_i^\sigma \tag{11b}$$

As is known, the Gibbs-Duhem equation

$$S^\alpha dT^\alpha - V^\alpha dP^\alpha + \sum_{i=1}^{n} N_i^\alpha d\mu_i^\alpha = 0 \tag{12a}$$

follows from Eqs. (10a) and (10b). Division by the total number of particles $N^\alpha = \sum_{i=1}^{n} N_i^\alpha$ gives

$$s^\alpha dT^\alpha - \upsilon^\alpha dP^\alpha + \sum_{i=1}^{n} x_i^\alpha d\mu_i^\alpha = 0 \tag{12b}$$

The similar equation follows from Eqs. (11a) and (11b):

$$Ad\sigma + S^\sigma dT^\sigma - V^{\alpha\sigma} dP^\alpha - V^{\beta\sigma} dP^\beta + \sum_{i=1}^{n} N_i^\sigma d\mu_i^\sigma = 0 \tag{13a}$$

Dividing this equation by the number of particles in the surface layer $N^\sigma = \sum_{i=1}^{n} N_i^\sigma$, we get

$$ad\sigma + s^\sigma dT^\sigma - \upsilon^{\alpha\sigma} dP^\alpha - \upsilon^{\beta\sigma} dP^\beta + \sum_{i=1}^{n} x_i^\sigma d\mu_i^\sigma = 0 \tag{13b}$$

where $a = A/N^\sigma$, $s^\sigma = S^\sigma/N^\sigma$, $\upsilon^{\alpha\sigma} = V^{\alpha\sigma}/N^\sigma$, $\upsilon^{\beta\sigma} = V^{\beta\sigma}/N^\sigma$, $x_i^\sigma = N_i^\sigma/N^\sigma$. It is seen that these specific quantities are the *mean* values for the surface layer, by definition.

From equation (13b) in the case of $n = 1$, we find an equation for $d\mu^\sigma$:

$$d\mu^\sigma = -s^\sigma dT^\sigma + \upsilon^{\alpha\sigma} dP^\alpha + \upsilon^{\beta\sigma} dP^\beta - ad\sigma \tag{14a}$$

Its multicomponent extension for $\mu_i^\sigma(T^\sigma, P^\alpha, P^\beta, \sigma, \mathbf{x}^\sigma)$ is

$$d\mu_i^\sigma = -s_i^\sigma dT^\sigma + \upsilon_i^{\alpha\sigma} dP^\alpha + \upsilon_i^{\beta\sigma} dP^\beta - a_i d\sigma + \sum_{j=1}^{n-1} \mu_{ij}^\sigma dx_j^\sigma, \quad \mu_{ij}^\sigma \equiv \left(\frac{\partial \mu_i^\sigma}{\partial x_j^\sigma}\right)_{T^\sigma, P^\alpha, P^\beta, \sigma, x_{k\neq j}^\sigma} \tag{14b}$$

where $s_i^\sigma$, etc. are the partial molecular quantities,

$$s_i^\sigma = \left(\frac{\partial S^\sigma}{\partial N_i^\sigma}\right)_{T^\sigma, P^\alpha, P^\beta, \sigma, N_{j\neq i}^\sigma}, \quad a_i = \left(\frac{\partial A}{\partial N_i^\sigma}\right)_{T^\sigma, P^\alpha, P^\beta, \sigma, N_{j\neq i}^\sigma}, \text{ etc.}$$

For a bulk phase,

$$d\mu_i^\alpha = -s_i^\alpha dT^\alpha + \upsilon_i^\alpha dP^\alpha + \sum_{j=1}^{n-1} \mu_{ij}^\alpha dx_j^\alpha, \quad \mu_{ij}^\alpha \equiv \left(\frac{\partial \mu_i^\alpha}{\partial x_j^\alpha}\right)_{T^\alpha, P^\alpha, x_{k\neq j}^\alpha} \tag{15}$$

Meaning the application of the foregoing equations to a *noncritical* nucleus (or to the corresponding DF), one remark concerning the concept of internal equilibrium of the surface layer should be made. The mechanical equilibrium is one of the components of internal equilibrium. It obviously takes place for an incompressible droplet of arbitrary size, but not for a noncritical bubble that expands or shrinks. However, the velocity of this macroscopic movement near the saddle point is very small and the above equations are assumed to be applicable to bubbles also.

## 2.3. Equation for the work

Before the appearance of the DF (state 1), the volume, entropy, and energy of our system are respectively $V_1$, $S_1$, and

$$E_1 = T^\beta S_1 - P^\beta V_1 + \sum_{i=1}^{n} \mu_i^\beta N_i^{tot} \tag{16}$$

In state 2 (the DF and ambient phase $\beta$), we have

$$V_2 = V^\alpha + V^{\alpha\sigma} + V^{\beta\sigma} + V^\beta, \quad S_2 = S^\alpha + S^\sigma + S^\beta, \quad N_i^{tot} = N_i^\alpha + N_i^\sigma + N_i^\beta, \quad E_2 = E^\alpha + E^\sigma + E^\beta \tag{17}$$

where $E^{\alpha(\beta)}$ and $E^\sigma$ are given by Eqs. (10b) and (11b).

Substituting differences $\Delta E = E_2 - E_1$, $\Delta S = S_2 - S_1$, and $\Delta V = V_2 - V_1$ in Eq. (9), we get after simple transformations

$$W = \left\{ \sum_{i=1}^{n} \left( \mu_i^\alpha - \mu_i^\beta \right) N_i^\alpha + \left( T^\alpha - T^\beta \right) S^\alpha - \left( P^\alpha - P^\beta \right) V^\alpha \right\}$$

$$+ \left\{ \sum_{i=1}^{n} \left( \mu_i^\sigma - \mu_i^\beta \right) N_i^\sigma + \left( T^\sigma - T^\beta \right) S^\sigma - \left( P^\sigma - P^\beta \right) V^{\alpha\sigma} + \sigma A \right\} \tag{18}$$

This is a basic equation for the following study.

The Gibbs free energy for different phases is as follows. For bulk phase $\alpha$,

$$G^\alpha = E^\alpha - T^\alpha S^\alpha + P^\alpha V^\alpha = \sum_{i=1}^{n} \mu_i^\alpha N_i^\alpha \tag{19}$$

and the same equation for phase $\beta$. For a curved surface layer [24],

$$G^\sigma = E^\sigma - T^\sigma S^\sigma + P^\alpha V^{\alpha\sigma} + P^\beta V^{\beta\sigma} - \sigma A = \sum_{i=1}^{n} \mu_i^\sigma N_i^\sigma \tag{20}$$

$$dG^\sigma = -S^\sigma dT^\sigma + V^{\alpha\sigma} dP^\alpha + V^{\beta\sigma} dP^\beta - A d\sigma + \sum_{i=1}^{n} \mu_i^\sigma dN_i^\sigma \tag{21}$$

The equality of the second mixed derivatives gives

$$\left( \frac{\partial \mu_i^\sigma}{\partial T^\sigma} \right)_{P^\alpha, P^\beta, \sigma, N_{j\neq i}^\sigma} = -\left( \frac{\partial S^\sigma}{\partial N_i^\sigma} \right)_{T^\sigma, P^\alpha, P^\beta, \sigma, N_{j\neq i}^\sigma} = -s_i^\sigma, \quad \left( \frac{\partial \mu_i^\sigma}{\partial P^\alpha} \right)_{T^\sigma, P^\beta, \sigma, N_{j\neq i}^\sigma} = \left( \frac{\partial V^{\alpha\sigma}}{\partial N_i^\sigma} \right)_{T^\sigma, P^\alpha, P^\beta, \sigma, N_{j\neq i}^\sigma} = v_i^{\alpha\sigma}, \text{ etc.}$$

$$\tag{22}$$

from where Eq. (14b) follows.

Thus, we have for states 1 and 2

$$G_1 = \sum_{i=1}^{n} \mu_i^{\beta} \left(N_i^{\alpha} + N_i^{\sigma} + N_i^{\beta}\right), \quad G_2 = G^{\alpha} + G^{\sigma} + G^{\beta} \tag{23}$$

and for difference $\Delta G = G_2 - G_1$,

$$\Delta G = \sum_{i=1}^{n} \left\{ \left(\mu_i^{\alpha} - \mu_i^{\beta}\right)N_i^{\alpha} + \left(\mu_i^{\sigma} - \mu_i^{\beta}\right)N_i^{\sigma} \right\} \tag{24}$$

The quantity $\Delta G$ is often employed in literature as a nucleation work in isothermal-isobaric conditions. Comparison of Eqs. (18) and (24) shows that $W \neq \Delta G$. Though phase $\beta$ is maintained at the isothermal-isobaric condition, the temperature and pressure of phase $\alpha$ differ from those of phase $\beta$, i.e. the isothermal-isobaric condition does not relate to the *whole* system (cf. also Ref. [33], § 20); the interface creates the difference in pressures and gives the contribution $\sigma A$ to $W$. For these reasons, the use of Gibbs' potential fails for calculating $W$ and Eq. (9) is employed here as the most relevant one. It is interesting that the macroscopic kinetic equations of nucleus evolution lead to the energy balance equation which consistently provides the constancy of $T^{\beta}$ and $P^{\beta}$, as it was revealed recently [13, 14].

## 3. Thermodynamics of a critical nucleus

### 3.1. The first differential of the work

Calculating the first differential of Eq. (18) at a fixed state of phase $\beta$ (this condition is denoted by subscript $\beta$) and with the use of Eqs. (12a) and (13a), we get

$$\left(dW\right)_{\beta} = \sum_{i=1}^{n} \left(\mu_i^{\alpha} - \mu_i^{\beta}\right)dN_i^{\alpha} + \sum_{i=1}^{n} \left(\mu_i^{\sigma} - \mu_i^{\beta}\right)dN_i^{\sigma} + \left(T^{\alpha} - T^{\beta}\right)dS^{\alpha} + \left(T^{\sigma} - T^{\beta}\right)dS^{\sigma}$$

$$-\left(P^{\alpha} - P^{\beta}\right)dV + \sigma dA, \qquad V \equiv V^{\alpha} + V^{\alpha\sigma} \tag{25}$$

As it was mentioned above, equation $\left(dW\right)_{\beta} = 0$ determines the saddle point of the considered system – the point of unstable equilibrium between the new-phase (critical) nucleus and phase $\beta$. From this equation and Eq. (25), the familiar conditions of equilibrium follow:

$$T^{\alpha} = T^{\sigma} = T^{\beta} \equiv T \tag{26a}$$

$$\mu_i^{\alpha} = \mu_i^{\sigma} = \mu_i^{\beta} = \mu_i \tag{26b}$$

$$P^{\alpha} - P^{\beta} = \frac{2\sigma}{R} \equiv P_L \tag{26c}$$

where $T$ and $\mu_i$ are the common temperature and chemical potential for the whole system in equilibrium. Eqs. (26a-c) determine the parameters of the critical nucleus via the known parameters of phase $\beta$.

It should be noted that Eqs. (26a-c) can be also obtained without the assumption of a fixed state of phase $\beta$. For this purpose, general Gibbs' condition of equilibrium

$$(dE_2)_{S_2,V_2,N_i^{tot}} = 0 \qquad (27)$$

is employed; all quantities included in this equation are given by Eq. (17). The differential $dE_2$ can be represented in the following form:

$$dE_2 = \sum_{i=1}^{n} \mu_i^\beta dN_i^{tot} + T^\beta dS_2 - P^\beta dV_2 + \sum_{i=1}^{n}\left(\mu_i^\alpha - \mu_i^\beta\right)dN_i^\alpha + \sum_{i=1}^{n}\left(\mu_i^\sigma - \mu_i^\beta\right)dN_i^\sigma$$

$$+ \left(T^\alpha - T^\beta\right)dS^\alpha + \left(T^\sigma - T^\beta\right)dS^\sigma - \left(P^\alpha - P^\beta\right)dV + \sigma dA \qquad (28)$$

From Eqs. (28) and (27), Eqs. (26a-c) follow.

It is seen that Eqs. (25) and (28), as well as Eq. (18), contain the volume $V$ of the region bounded by the surface of tension; Eq. (26c) also relates to this surface. It is natural to consider this region as a *new-phase nucleus*, in accordance with conventional treatment. At then same time, the DF volume is larger by $V^{\beta\sigma}$ (Fig. 1); hence, the nucleus is the DF core. Substitution of Eqs. (26a-c) in Eq. (18) yields Gibbs' formula for the work of critical nucleus formation

$$W_* = \frac{1}{3}\sigma A_* \qquad (29)$$

If the change in the considered three-phase system state is performed while maintaining the equilibrium between all the phases, then equation

$$d\mu_i^\alpha = d\mu_i^\sigma = d\mu_i^\beta = d\mu_i \qquad (30)$$

holds in addition to Eq. (26b). From this equation, different relations between the parameters of coexisting phases can be derived. Such relations in the form of differential equations, in particular, equations for various dependences of the surface tension on parameters of coexisting phases, were derived by Rusanov [24] in the general case of a multicomponent system; the partial derivatives of the molar Gibbs' potential with respect to the molar fractions, $\partial g / \partial x_i$, were employed for this purpose (the conditions of equilibrium for $\partial g / \partial x_i$ are the same as Eqs. (26b) and (30)) . Below, only some relations relevant for the present study are considered for a single-component system together with different derivation: explicit Eqs. (14a) and (15) for the differentials of chemical potentials and the particular cases of Eq. (30),

$$d\mu_i^\alpha = d\mu_i^\beta \qquad (31a)$$

$$d\mu_i^\alpha = d\mu_i^\sigma \qquad (31b)$$

$$d\mu_i^\sigma = d\mu_i^\beta \qquad (31c)$$

are used here instead of $\partial g / \partial x_i$.

## 3.2. Dependence of surface tension on radius

Eq. (31a) at constant $T$ is

$$v^\alpha dP^\alpha = v^\beta dP^\beta \qquad (32)$$

Substituting here

$$dP^\alpha = dP^\beta + dP_L \qquad (33)$$

according to Eq. (26c), we obtain an equation for the equilibrium vapor pressure $P^\beta$ for the droplet ($\alpha$) of radius $R$,

$$\frac{dP^\beta}{dP_L} = \frac{v^\alpha}{v^\beta - v^\alpha} \qquad (34a)$$

In the CNT approximation, the presence of phase $\sigma$ is neglected, so that this equation relates to *adjacent* phases $\alpha$ and $\beta$. The boundary condition to it is the saturation pressure over the planar interface

$$P^\beta(R = \infty) \equiv P_S(T) = C\,e^{-\frac{q_{(\alpha\beta)}}{kT}} \qquad (34b)$$

obtained from the Clapeyron-Clausius equation which is also an equation for adjacent phases; $q_{(\alpha\beta)} \equiv T^\beta(s^\beta - s^\alpha)$ is the heat of transition $\beta \to \alpha$, or the CNT heat of vaporization; $C$ is the constant. Integration of Eq. (34a) for an incompressible liquid ($v^\alpha$ does not depend on $P_L$) from $P_L = 0$ ($R = \infty$) to the current $P_L$, for ideal vapor, $v^\beta = kT / P^\beta$, and far from the critical point, $v^\alpha << v^\beta$, gives the familiar Kelvin equation

$$P^\beta = P_S(T)\,e^{\frac{v^\alpha P_L}{kT}} = P_S(T)\,e^{\frac{2v^\alpha \sigma}{kTR}} \qquad (35)$$

with constant $\sigma$.

For deriving the isothermal dependence of the surface tension on radius $\sigma(R)_T$, Eq. (31c) is employed with Eq. (14a):

$$-s^\sigma dT + v^{\alpha\sigma} dP^\alpha + v^{\beta\sigma} dP^\beta - a d\sigma = -s^\beta dT + v^\beta dP^\beta \qquad (36)$$

Considering this equation at constant temperature, $dT = 0$, and utilizing Eqs. (33) and (34a), we get

$$a(d\sigma)_T = \omega' dP_L = -\omega dP_L \qquad (37)$$

where

$$\omega \equiv v^{\beta\sigma} - v^\beta \frac{v^\sigma - v^\alpha}{v^\beta - v^\alpha} = -\omega'\ , \quad \omega' = v^{\alpha\sigma} - v^\alpha \frac{v^\sigma - v^\beta}{v^\alpha - v^\beta}, \quad v^\sigma = v^{\alpha\sigma} + v^{\beta\sigma} \qquad (38)$$

The desired equation for $\sigma(R)_T$ is obtained by substituting here

$$dP_L = \frac{2}{R} d\sigma - \frac{2\sigma}{R^2} dR \tag{39}$$

The result is [24]

$$\frac{1}{\sigma}\left(\frac{d\sigma}{dR}\right)_T = \frac{2\tilde{\delta}/R^2}{1+2\tilde{\delta}/R} \equiv \phi(R), \quad \tilde{\delta} \equiv \frac{\omega}{a} \tag{40}$$

Integration of this equation gives

$$\sigma(R)_T = \sigma_\infty(T)\Phi(R), \quad \Phi(R) \equiv \exp\left[-\int_R^\infty \phi(R')dR'\right] \tag{41}$$

where $\sigma_\infty(T)$ is the surface tension of the planar interface. If the quantity $\tilde{\delta}$ does not depend on $R$, the integral is easily calculated giving

$$\sigma(R)_T = \frac{\sigma_\infty(T)}{1+2\tilde{\delta}/R} \tag{42}$$

The asymptotics of the function $\phi(R)$ are

$$\phi(R) = \begin{cases} 2\tilde{\delta}/R^2, & R \to \infty \\ 1/R, & R \to 0 \end{cases} \tag{43}$$

Accordingly, at constant $\tilde{\delta}$,

$$\sigma(R)_T = \begin{cases} \sigma_\infty(T)(1-2\tilde{\delta}/R), & R \to \infty \\ K(T)R, & R \to 0 \end{cases} \tag{44}$$

The linear asymptotics of $\sigma(R)_T$ at $R \to 0$ was found in Ref. [24] from a special analysis.

The superficial number of particles $N_\Sigma$ and adsorption $\Gamma$ in Gibbs' approach are

$$N_\Sigma = N^\sigma - \rho^\alpha V^{\alpha\sigma} - \rho^\beta V^{\beta\sigma}, \quad \Gamma = N_\Sigma/A \tag{45}$$

It is easy to express the quantity $\tilde{\delta}$ in terms of $\Gamma$ as follows [24]:

$$\tilde{\delta} = \frac{\Gamma}{\rho^\alpha - \rho^\beta}, \tag{46}$$

Thus, Eq. (40) goes to that derived by Gibbs,

$$\frac{1}{\sigma}\left(\frac{d\sigma}{dR}\right)_T = \frac{(2/R^2)(\Gamma/(\rho^\alpha - \rho^\beta))}{1+(2/R)(\Gamma/(\rho^\alpha - \rho^\beta))} \tag{47}$$

Tolman [35] found that

$$\frac{\Gamma}{\rho^\alpha - \rho^\beta} = \tilde{\delta} = \delta\left[1 + \frac{\delta}{R} + \frac{1}{3}\left(\frac{\delta}{R}\right)^2\right] \tag{48}$$

where $\delta = R_{em} - R$ is the spacing between the equimolecular surface corresponding to $\Gamma = 0$ and the surface of tension (Fig. 1). Substitution of Eq. (48) in Eq. (47) gives the familiar Tolman equation; Eq. (42) or the first Eq. (44) with $\tilde{\delta} = \delta$ is Tolman's approximation for $\sigma(R)_T$ in the case of $(\delta/R) << 1$ and $\delta = const \equiv \delta_\infty$.

For the vapor-liquid ($\beta$-$\alpha$) interface far from the critical point, we can assume $v^\alpha << v^\beta$ and $v^\sigma << v^\beta$. Eq. (38) is then simplified as

$$\omega = a\tilde{\delta} \approx v^\alpha - v^{\alpha\sigma} \qquad (49)$$

It should be noted that the derivation of Eq. (37) for a droplet and a bubble gives $a(d\sigma)_T = \omega' dP_L$ and $a(d\sigma)_T = \omega dP_L$, respectively. However, the superscripts $\alpha$ and $\beta$ in Eq. (38) for $\omega'$ relate to liquid and vapor, respectively, and *vice versa* in $\omega$. Replacing then $\alpha \leftrightarrow \beta$ in $\omega$, we arrive at $\omega'$. Thus, Eqs. (37) and (40) hold both for droplets and for bubbles and the quantity $\omega$, Eq. (49), does not change sign for bubbles, as is stated in Ref. [24]. The only difference between droplets and bubbles may be due to different dependences $v^{\alpha\sigma}(R)$. The assumption of opposite signs for $\omega$ would result in the conclusion that $\delta_\infty = 0$, since both a droplet and a bubble approach the same planar interface at $R \to \infty$.

### 3.3. Extension of Kelvin's formula to small radii

The dependence $P^\beta(R)$, Eq. (35), was derived above for adjacent phases $\alpha$ and $\beta$. In the case of small droplet radii, especially when phase $\alpha$ is absent, it is not suitable more and has to be replaced by the equation following from the equilibrium between adjacent phases $\sigma$ and $\beta$. Eq. (36) at constant $T$ with the use of Eqs. (33) acquires the form

$$(v^\beta - v^\sigma)dP^\beta = v^{\alpha\sigma}dP_L - ad\sigma = \left\{ v^{\alpha\sigma}(R)\frac{dP_L}{dR} - a(R)\frac{d\sigma}{dR} \right\}dR \qquad (50)$$

In the ideal gas model, the condition $v^\sigma << v^\beta$ holds. Employing Eqs. (39)-(41) and then integrating this equation over $R$, as before, we get

$$\ln\frac{P^\beta}{P_S^\sigma} = \frac{\sigma_\infty(T)\Psi(R)}{kT}, \quad \Psi(R) \equiv \int_R^\infty \left\{ \frac{2v^{\alpha\sigma}(R)}{R^2} + \left[ a(R) - \frac{2v^{\alpha\sigma}(R)}{R} \right]\phi(R) \right\}\Phi(R)dR, \quad P_S^\sigma(T) = C^\sigma e^{-\frac{q_{(\sigma\beta)}}{kT}} \qquad (51)$$

where $q_{(\sigma\beta)} \equiv T^\beta(s^\beta - s^\sigma)$ is the heat of transition $\beta \to \sigma$. Evidently, just this quantity should be identified with the heat of vaporization, if the latter is determined from the saturation pressure $P_S$.

If $v^{\alpha\sigma}$ and $a$ are assumed not depending on $R$ for sufficiently large droplets, then Eq. (50) results in the following explicit equation:

$$P^\beta = P_S^\sigma(T)e^{\frac{v^{\alpha\sigma}P_L - a(\sigma(R)_T - \sigma_\infty)}{kT}} \qquad (52)$$

The ideal gas model may be unsatisfactory at high supersaturations (small critical radii). Assuming the fulfillment of the condition $v^\sigma << v^\beta$, as before, we have instead of Eq. (51)

$$\int\limits_{P_S}^{P^\beta} \frac{dP^\beta}{\rho^\beta(P^\beta)} = \sigma_\infty(T)\Psi(R) \tag{53}$$

where $\rho^\beta(P^\beta)$ is the equation of state of the vapor; hereafter the pressure $P_S$ is written without the superscript $\sigma$. In terms of the variable $z = \ln(P^\beta / P_S)$, Eq. (53) has the form

$$F(z) \equiv \frac{P_S}{kT} \int\limits_0^z \frac{e^x \, dx}{\rho^\beta(P_S \, e^x)} = \frac{\sigma_\infty(T)\Psi(R)}{kT} \tag{54}$$

The Van der Waals equation

$$P(\rho) = \frac{\rho kT}{1 - b_W \rho} - a_W \rho^2, \quad b_W = \frac{kT_c}{8P_c} 0.5, \quad a_W = \frac{27}{8} kT_c b_W \tag{55}$$

is used here as the equation of state; $T_c$ and $P_c$ are the critical temperature and pressure for water. The parameter $b_W$ is multiplied by the factor 0.5 in order to get a more realistic liquid density 33.2 nm$^{-3}$ which is close to water density; without this factor, the density is half.

Solving cubic with respect to $\rho$ Eq. (55), we find three roots $\rho^{(i)}(P)$ corresponding to vapor, liquid, and nonexistent (unstable) phases (Fig. 2a). The vapor curve ends at the pressure $P_m$ =62.8 bar which is the limiting (spinodal) pressure for the vapor. Hence, the limiting supersaturation at $T = 20 \ ^\circ C$ is $P_m / P_S = 2.7 \times 10^3$ and $z_m = 7.9$; $\rho^\alpha / \rho^\beta = 0.09$ and $\upsilon^\beta / \upsilon^\alpha = 10.6$ at this point. The function $F(z)$ is plotted in Fig. 2b together with the corresponding function $F(z) = z$ for ideal vapor. It is seen that both the plots are practically coincide up to the highest supersaturations. It was checked also that the modification of the function $F(z)$ by taking into account $\upsilon^\sigma$, Eq. (50), does not change Fig. 2b. Thus, the ideal gas model and Eq. (51) are good for calculating the critical radius for the given supersaturation (or vice versa) in the practical region of nucleation.

It should be noted that Eq. (51) cannot be obtained within Gibbs' approach, since the consideration of a *real* surface layer is crucial here.

### 3.4. Dependence of surface tension on temperature

The set of Eqs. (31b) and (31c) with account for Eq. (14a) becomes as follows:

$$\begin{cases} a d\sigma = (s^\alpha - s^\sigma)dT + (\upsilon^{\alpha\sigma} - \upsilon^\alpha)dP^\alpha + \upsilon^{\beta\sigma}dP^\beta \\ a d\sigma = (s^\beta - s^\sigma)dT + \upsilon^{\alpha\sigma}dP^\alpha + (\upsilon^{\beta\sigma} - \upsilon^\beta)dP^\beta \end{cases} \tag{56}$$

Utilizing Eq. (33) and excluding $dP^\beta$ from these equations as well as employing Eq. (39), we get the following resulting equation:

$$\left(a + \frac{2\omega}{R}\right)d\sigma = \left[\frac{s^\beta(\upsilon^\sigma - \upsilon^\alpha) + s^\alpha(\upsilon^\beta - \upsilon^\sigma)}{\upsilon^\beta - \upsilon^\alpha} - s^\sigma\right]dT + \frac{2\omega\sigma}{R^2}\,dR \tag{57}$$

It is seen that Eq. (40) is again obtained from here at constant $T$.

So, an equation for the dependence of surface tension on temperature at constant radius is

$$\left(\frac{d\sigma}{dT}\right)_R = \left(a + \frac{2\omega}{R}\right)^{-1}\left[\frac{s^\beta(v^\sigma - v^\alpha) + s^\alpha(v^\beta - v^\sigma)}{v^\beta - v^\alpha} - s^\sigma\right] \qquad (58)$$

This equation is also simplified for the vapor-liquid interface far from the critical point. In addition, we assume that the surface layer state in a one-component system is intermediate with respect to the states of coexisting bulk phases: $s^\alpha < s^\sigma < s^\beta$, $v^\alpha < v^\sigma < v^\beta$. So,

$$\left(a + \frac{2\omega}{R}\right)\left(\frac{d\sigma}{dT}\right)_R \approx s^\alpha - s^\sigma = -\frac{q_{(\alpha\sigma)}}{T} \qquad (59)$$

where $q_{(\alpha\sigma)} = T(s^\sigma - s^\alpha)$ is the heat of transition $\sigma \rightarrow \alpha$. Apparently, $s^\sigma$ depends on $R$, hence, $q_{(\alpha\sigma)}$ is a function of $R$ also.

For the planar interface,

$$a_\infty \frac{d\sigma_\infty}{dT} \approx s^\alpha - s^\sigma < 0 \qquad (60)$$

which is the experimentally known fact: the surface tension of liquids decreases with increasing temperature. As shown below, the term $(a + 2\omega/R)$ is positive for droplets in the nucleation region. Hence, the derivative $d\sigma/dT$ is negative for small droplets also.

Eq. (57) and following from it Eqs. (40) and (58) are obtained from the condition of equilibrium between all phases - $\alpha$, $\sigma$, and $\beta$; i.e. the change in state, Eq. (30), concerns all phases, including phase $\beta$. At the same time, it is of interest to consider the equilibrium only between phases $\alpha$ and $\sigma$ at a fixed state of phase $\beta$, which is important for the present study. This is the *DF internal equilibrium* which is usually assumed in the theory of nucleation; the DF corresponding to a noncritical nucleus is not in equilibrium with phase $\beta$.

As mentioned above, the condition of mechanical equilibrium, Eqs. (26c) and (33), holds for a noncritical droplet also. From equation (31b) together with Eq. (33) and $dP^\beta = 0$, we get

$$ad\sigma = (s^\alpha - s^\sigma)dT - (v^\alpha - v^{\alpha\sigma})dP_L \qquad (61a)$$

and with account for Eq. (39),

$$\left(a + \frac{2\omega}{R}\right)d\sigma = (s^\alpha - s^\sigma)dT + \frac{2\omega\sigma}{R^2}dR \qquad (61b)$$

where $T \equiv T^\alpha = T^\sigma$ and $\omega \approx v^\alpha - v^{\alpha\sigma}$, Eq. (49). Thus, we have an interesting result: Eq. (61b) is nothing but Eq. (57) for a droplet in the vapor far from the critical point. In other words, the condition of DF internal equilibrium at a fixed state of the ambient phase gives for a droplet in vapor the same dependence of the surface tension on radius and temperature, $\sigma = \sigma(R,T)_\beta$, as the condition of total equilibrium; i.e. considering the internal equilibrium only, we can use Eqs. (40) and (59) obtained from the condition of total equilibrium (where $T$ is the DF temperature now). This important fact will be employed later.

It follows from Eq. (61b) that the surface tension depends on $R$ and $T$ due to the difference in physical properties of the surface layer and bulk phase $\alpha$: the dependence on $R$ is due to nonzero $\omega$, i.e. the difference in specific volumes $v^{\alpha}$ and $v^{\alpha\sigma}$, whereas the dependence on $T$ arises from the difference in entropies.

### 3.5. Numerical estimates

As is seen from the foregoing, the knowledge of the dependence of Tolman's length on radius is necessary for calculating the thermodynamic characteristics of nucleation. Unfortunately, this dependence is unknown as yet. There is disagreement in literature even about the asymptotic value $\delta_{\infty}$ of this quantity. Tolman [36] was the first who estimated $\delta_{\infty}$ for water and obtained for it the positive value about 0.1 nm = $0.33\,d$, where $d = 0.3$ nm is the molecular diameter. Rusanov [24] estimated the quantity $\omega$ for the planar interface, Eq. (49), as positive also. The assessment was made for $CCl_4$ on the basis of the formula for the surface layer thickness derived within the statistical mechanical approach. However, later, the dependence $\delta(R_{em})$ was calculated numerically for water clusters [37] and it was shown that $\delta$ changes from positive to negative values with increasing $R_{em}$. Both negative [38-44], and positive [45-49] values of $\delta_{\infty}$ are reported in literature.

In order to make numerical estimates, some simple functions for $\delta(R)$ are used here which take into account the different mentioned possibilities for $\delta_{\infty}$:

$$\text{(I)} \quad \delta(R) = l_1 \exp(-R/l_2) + \delta_{\infty}, \quad l_1 = 8.9452 \times 10^{-2}, \; l_2 = 10, \; \delta_{\infty} = -0.045 \quad \text{(62a)}$$

$$\text{(I')} \quad \delta(R) = l_1 \exp(-R/l_2), \quad l_1 = 4.2595 \times 10^{-2}, \; l_2 = 10 \quad (\delta_{\infty} = 0) \quad \text{(62b)}$$

$$\text{(II)} \quad \delta(R) = l_1 \left[1 - \exp(-l_2/R)\right] + \delta_{\infty}, \quad l_1 = 8.6664 \times 10^{-2}, \; l_2 = 10, \; \delta_{\infty} = -0.045 \quad \text{(62c)}$$

$$\text{(III)} \quad \delta(R) = \delta_0 > 0, \quad \delta_0 = 4.0955 \times 10^{-2} \quad \text{(62d)}$$

where all the parameters are in nm.

It was found [38-42] that $\delta_{\infty}$ is within $-(0.1 - 0.2)\,d$, hence, the value $\delta_{\infty} = -0.15\,d$ for water was taken. The values of $l_1$ and $\delta_0$ were fitted in order to get the limiting supersaturation $z_m$ mentioned above; at these values, the equation

$$F(z_m) = \frac{\sigma_{\infty}(T)\Psi(0)}{kT} \quad \text{(63a)}$$

is obeyed. It should be noted that $\Psi(0)$ rather weakly depends on $l_2$, but very sensitive to $l_1$ (which is reflected by the great number of decimal digits in its representation). The parameters $l_1$ and $\delta_{\infty}$ determine $\delta(0)$, $\delta(0) = l_1 - \delta_\infty$, so that this quantity is determined by the limiting supersaturation. These functions and the corresponding dependences $\sigma(R)$ for water are plotted in Figs. 3a, b. The maxima of curves I and II are pronounced very weekly and located far beyond the region of nucleation. All the curves $\sigma(R)$ have the same linear asymptotics at $R \to 0$, Eq. (44), with $K = 5406$ bar; $K$ also relates to the limiting supersaturation. From Eq. (34a) at $\upsilon^\alpha << \upsilon^\beta$, one follows that

$$P_L(P^\beta) = \rho^\alpha kTF(z) \tag{63b}$$

$P_L \to 2K$ at $z \to z_m$, hence

$$2K = \rho^\alpha kTF(z_m) \tag{63c}$$

This equation gives $2K = 1.081 \times 10^4$ bar, in accordance with $\sigma(R)$ calculations.

Dashed curve in Fig. 3b corresponds to function I'. It is noteworthy that it is very close to curve II. As is seen from Fig. 3a, quite different functions I' and II have only two common features: (i) close values of $\delta(0)$ and (ii) both decrease with increasing $R$ (though approach the different values of $\delta_\infty$, 0 and $-0.045$). Function III has about the same value of $\delta(0)$, however, it is constant. Function I is decreasing, however, its $\delta(0)$ somewhat differs. This analysis suggests that (i) just the quantity $\delta(0)$ is significant in the theory; it is much more significant, than $\delta_\infty$ (also, the asymptotics given by Eq. (42) is of no interest for nucleation, since it is valid far beyond the nucleation region). (ii) The real dependence $\delta(R)$ begins with small positive value $\delta(0)$ and decreases with increasing $R$ approaching, probably, a negative value of $\delta_\infty$. It was checked that the case with negative $\delta(0)$ does not lead to reasonable physical results; it does not allow obtaining the dependence $\sigma(R)$ at small $R$.

The suggestion that the curvature effect (the dependence of surface tension on curvature) is determined by $\delta(0)$ is supported by Figs. 3c, d. In Fig. 3c, three versions of function II with the same $\delta(0) = 0.0417$ nm and different values of $\delta_\infty$ are shown:

(1) $\delta_\infty = -0.06$, $l_2 = 11.16$; (2) $\delta_\infty = -0.03$, $l_2 = 7.845$; (3) $\delta_\infty = 0$, $l_2 = 4.53$

all these lengths are in nm. For the given values of $\delta(0)$ and $\delta_\infty$, the values of $l_2$ were fitted to obey Eq. (63a). The dependences $\sigma_T(R)$ corresponding to these functions are shown in Fig. 3d; they coincide in the nucleation region, e.g. $\sigma_{(1)} = \sigma_{(2)} = \sigma_{(3)} = 65.98$ erg/cm$^2$ at $R = 0.7$ nm.

As noted above, the quantity $\delta(0)$ is determined by the limiting supersaturation. If we invert the function $\delta(R)$ and consider $\delta$ as an independent variable and $R = R(\delta)$, then $\delta(0)$ can be found from Eq. (63a).

The parts $(\alpha\sigma)$ and $(\beta\sigma)$ of the surface layer can be called respectively "liquid-like" and "vapor-like". Let $\rho^{\alpha\sigma}$ and $\rho^{\beta\sigma}$ are the mean densities of these parts; it should be recalled that $\upsilon^{\alpha\sigma} = V^{\alpha\sigma}/N^\sigma \neq 1/\rho^{\alpha\sigma}$ and similarly for $\rho^{\beta\sigma}$. Then

$$N^\sigma = N^{\alpha\sigma} + N^{\beta\sigma}, \quad N^{\alpha\sigma} = \rho^{\alpha\sigma} V^{\alpha\sigma}, \quad N^{\beta\sigma} = \rho^{\beta\sigma} V^{\beta\sigma} \tag{64}$$

By definition, $N_\Sigma = 0$ for the equimolecular surface; thus, Eq. (45) has the form

$$\begin{cases} \rho^{\alpha\sigma} V^{\alpha\sigma} + \rho^{\beta\sigma} V^{\beta\sigma} - \rho^\alpha (V^{\alpha\sigma} + V_\delta) - \rho^\beta (V^{\beta\sigma} - V_\delta) = 0, & \delta > 0 \\ \rho^{\alpha\sigma} V^{\alpha\sigma} + \rho^{\beta\sigma} V^{\beta\sigma} - \rho^\alpha (V^{\alpha\sigma} - V_\delta) - \rho^\beta (V^{\beta\sigma} + V_\delta) = 0, & \delta < 0 \end{cases} \tag{65}$$

where $V_\delta$ is the volume of the layer of width $\delta$ (Fig. 1). Such separation into two cases concerns only functions I and II with $\delta_\infty < 0$. Let $R_0$ denotes the distance at which $\delta$ changes the sign, $\delta(R_0) = 0$.

Denoting $\eta \equiv (\rho^\alpha - \rho^{\alpha\sigma})/\rho^\alpha$ and employing the conditions $\rho^\alpha >> \rho^\beta$, $\rho^{\beta\sigma} >> \rho^\beta$, we get from Eq. (65)

$$V^{\beta\sigma} = \begin{cases} (\rho^\alpha/\rho^{\beta\sigma})(\eta V^{\alpha\sigma} + V_\delta), & \delta > 0 \\ (\rho^\alpha/\rho^{\beta\sigma})(\eta V^{\alpha\sigma} - V_\delta), & \delta < 0 \end{cases} \tag{66}$$

The natural condition $V^{\beta\sigma} > 0$ requires

$$\eta > \eta_0 \equiv \left( \frac{V_\delta}{V - V^\alpha} \right)_{\max} \tag{67}$$

where $V - V^\alpha = V^{\alpha\sigma}$. Thus, the case of $\delta_\infty < 0$ imposes the restriction on the values of density $\rho^{\alpha\sigma}$. The quantity $\eta_0$ is sensitive to the parameter $l_2$, so this parameter was chosen to make $\eta_0$ as less as possible. Calculations were performed for the case, when phase $\alpha$ is absent, hence, $V^{\alpha\sigma} = V = 4\pi R^3/3$. For this case, $\eta_0 = 5\times10^{-3}$ and $5.7\times10^{-4}$ for functions I and II; $R_0$ is respectively 6.9 and 13.7 nm, which is far beyond the region of nucleation (~ 1 nm).

With the use of Eq. (66), one obtains

$$N^\sigma(R) = (1-\eta)\rho^\alpha V^{\alpha\sigma} + \rho^{\beta\sigma} V^{\beta\sigma} = \rho^\alpha V_{em}(R), \quad V_{em}(R) \equiv V^{\alpha\sigma} \pm V_\delta(R), \quad V_\delta(R) = 4\pi R^2 \left| \widetilde{\delta}(R) \right| \tag{68}$$

So, the important quantity $N^\sigma$ is determined only by the bulk density $\rho^\alpha$ and Tolman's length and does not depend on $V^{\beta\sigma}$. The volume $V^{\beta\sigma}$ itself depends, in addition, on density $\rho^{\beta\sigma}$. The Laplace pressure for a critical droplet is about $1.5\times10^3$ bar, so the bulk density was taken with account for the compressibility, $\rho^\alpha = 1.053\times\rho_0^\alpha = 35.23$ nm$^{-3}$; also, it was put $\rho^{\beta\sigma} = 0.5\,\rho^\alpha$ and $\eta = 0.03$. The numbers $N^{\alpha\sigma}$, $N^{\beta\sigma}$, and $N^\sigma$ for all the functions are obtained close to each other: $N^{\alpha\sigma} = 49.1$, $N^{\beta\sigma} = 10.3 - 11.1$, and $N^\sigma = 59.4 - 60.2$ for the critical radius $R_* = 0.7$ nm. For comparison, CNT gives (without account for compressibility) $N_* = 48.1$.

Having $N^\sigma(R)$, it is possible to calculate

$$a(R) = \frac{4\pi R^2}{N^\sigma(R)}, \quad v^{\alpha\sigma}(R) = \frac{4\pi R^3}{3 N^\sigma(R)}, \quad v^{\beta\sigma}(R) = \frac{V^{\beta\sigma}(R)}{N^\sigma(R)}, \quad v^\sigma(R) = v^{\alpha\sigma}(R) + v^{\beta\sigma}(R) \tag{69}$$

The plots of these dependences have the similar form for all the above functions $\delta(R)$ and are shown in Fig. 4 for function I. The functions $\omega(R)$ according to Eqs. (38) and (49) coincide up to $R = 0$. The dependence $v^{\beta}(R)$ was calculated with account for the vapor supersaturation for the given $R$, according to Eq. (51); $v^{\beta}(0)/v^{\sigma}(0) = 15.5$.

# 4. Thermodynamics of a noncritical nucleus

All equations obtained in the previous Section except the equations for the DF internal equilibrium relate to a *critical* nucleus located at the saddle point of the work; they were derived from the condition of equilibrium between all three phases. It should be noted that these equations are valid independently of the actual presence of phase $\alpha$. If the nucleus is so small that bulk phase $\alpha$ is absent, the latter serves as a *reference phase*, in accordance with original Gibbs' treatment [23]; its parameters are uniquely determined by Eqs. (26a-c). Thermodynamic description makes sense for such small nuclei as well. The surface layer does not vanish at $R \to 0$ (Fig. 1), i.e. the DF (or the region of inhomogeneity) remains, which agrees with Gibbs' assumption of a small finite inhomogeneity remaining when the surface of tension vanishes; Eq. (66) gives $V^{\beta\sigma}(0) = (\rho^{\alpha}/\rho^{\beta\sigma})V_{\delta}(0)$. Therefore, the thermodynamic description of the system consisting of the DF and ambient phase is valid [24].

Below, two cases when there is no equilibrium between all phases, or the fluctuations, are considered : (i) fluctuations of surface layer parameters, when the $\sigma$-phase state deviates from the equilibrium with phases $\alpha$ and $\beta$; (ii) the DF corresponding to a noncritical nucleus.

## 4.1. The second differential of the work

Taking the differential of Eq. (25) at the fixed state of phase $\beta$, we obtain the second differential of the work. In view of the equality

$$\left[\sigma d^2 A - (P^{\alpha} - P^{\beta})d^2 V\right]_* = -\frac{P_L^*}{3V_*}(dV)^2 \tag{70}$$

it has the following form:

$$(d^2 W)_{\beta}^* = -\frac{P_L^*}{3V_*}(dV)^2 + \left\{\sum_{i=1}^{n} d\mu_i^{\alpha} dN_i^{\alpha} + dT^{\alpha} dS^{\alpha} - dP^{\alpha} dV^{\alpha}\right\}_*$$

$$+ \left\{\sum_{i=1}^{n} d\mu_i^{\sigma} dN_i^{\sigma} + dT^{\sigma} dS^{\sigma} - dP^{\alpha} dV^{\alpha\sigma} + d\sigma dA\right\}_* \equiv -\frac{P_L^*}{3V_*}(dV)^2 + H^{\alpha} + H^{\sigma} \tag{71}$$

where $H^{\alpha}$ and $H^{\sigma}$ are the positive definite quadratic forms of stable variables for phases $\alpha$ and $\sigma$ which are placed in braces. So, equation for the second differential naturally splits into two parts corresponding to the bulk and surface phases.

Using Eqs. (14b) and (15), as well as

$$S^{\alpha(\sigma)} = \sum_{i=1}^{n} s_i^{\alpha(\sigma)} N_i^{\alpha(\sigma)} , \quad V^{\alpha} = \sum_{i=1}^{n} \upsilon_i^{\alpha} N_i^{\alpha} , \quad V^{\alpha\sigma} = \sum_{i=1}^{n} \upsilon_i^{\alpha\sigma} N_i^{\sigma} , \quad A = \sum_{i=1}^{n} a_i N_i^{\sigma} \tag{72}$$

the quadratic forms can be transformed as

$$H^{\alpha} = \sum_{i=1}^{n} \left\{ N_i^{\alpha} \left[ ds_i^{\alpha} dT^{\alpha} - d\upsilon_i^{\alpha} dP^{\alpha} \right] + dN_i^{\alpha} \sum_{j=1}^{n-1} \mu_{ij}^{\alpha} dx_j^{\alpha} \right\}_* \tag{73a}$$

$$H^{\sigma} = \sum_{i=1}^{n} \left\{ N_i^{\sigma} \left[ ds_i^{\sigma} dT^{\sigma} - d\upsilon_i^{\alpha\sigma} dP^{\alpha} + da_i d\sigma \right] + dN_i^{\sigma} \sum_{j=1}^{n-1} \mu_{ij}^{\sigma} dx_j^{\sigma} \right\}_* \tag{73b}$$

For a single component system,

$$H^{\alpha} = N_*^{\alpha} \left[ ds^{\alpha} dT^{\alpha} - d\upsilon^{\alpha} dP^{\alpha} \right]_* \tag{74a}$$

$$H^{\sigma} = N_*^{\sigma} \left[ ds^{\sigma} dT^{\sigma} - d\upsilon^{\alpha\sigma} dP^{\alpha} + dad\sigma \right]_* \tag{74b}$$

Knowledge of the second differential makes it possible to get an explicit equation for the work of a near-critical nucleus formation according to Eq. (3) with subsequent numerical evaluation. Differentials in Eqs. (71)–(74) denote small deviations of the nucleus parameters from their equilibrium (saddle-point) values: $dV = V - V_*$, $dT^{\alpha} = T^{\alpha} - T^{\beta}$, etc. Comparison of Eqs. (71) and (74a, b) with Eqs. (4a) and (4b) derived within Gibbs' approach shows certain similarity between them. The advantage of Eq. (74b) in comparison with Eq. (4b) is in that it contains real physical parameters of the surface layer rather than superficial quantities. This fact allows calculating this quadratic form with the use of some approximations (which is demonstrated below) and making then numerical estimates of surface effects.

## 4.2. Fluctuations of surface layer

Eqs. (73b) and (74b) are basic for calculating the fluctuations of surface layer parameters (at $R = R_*$ and $N^{\sigma} = N_*^{\sigma}$). The single component case is considered here for simplicity. Two sets of independent variables for the quadratic form $H^{\sigma}$, Eq. (74b), are considered below. The first set is ($s^{\sigma}$, $P^{\alpha}$, $\sigma$), accordingly, $T^{\sigma}$, $\upsilon^{\alpha\sigma}$, and $a$ are functions of these variables.

An equation for the surface layer enthalpy per one molecule is [24]

$$h^{\sigma} = e^{\sigma} + P^{\alpha} \upsilon^{\alpha\sigma} + P^{\beta} \upsilon^{\beta\sigma} - \sigma a = Ts^{\sigma} + \mu^{\sigma}, \quad e^{\sigma} \equiv E^{\sigma} / N^{\sigma} \tag{75a}$$

from where, with account for Eq. (14a),

$$dh^{\sigma} = T^{\sigma} ds^{\sigma} + \upsilon^{\alpha\sigma} dP^{\alpha} + \upsilon^{\beta\sigma} dP^{\beta} - a d\sigma \tag{75b}$$

The term with $dP^{\beta}$ can be ignored in this equation, since the $\beta$-phase state is assumed to be fixed. From equality of the second mixed derivatives, the following relations hold:

$$\left( \frac{\partial \upsilon^{\alpha\sigma}}{\partial s^{\sigma}} \right)_{P^{\alpha}, \sigma} = \left( \frac{\partial T^{\sigma}}{\partial P^{\alpha}} \right)_{s^{\sigma}, \sigma} , \quad \left( \frac{\partial \upsilon^{\alpha\sigma}}{\partial \sigma} \right)_{s^{\sigma}, P^{\alpha}} = -\left( \frac{\partial a}{\partial P^{\alpha}} \right)_{s^{\sigma}, \sigma} , \quad \left( \frac{\partial T^{\sigma}}{\partial \sigma} \right)_{s^{\sigma}, P^{\alpha}} = -\left( \frac{\partial a}{\partial s^{\sigma}} \right)_{P^{\alpha}, \sigma} \tag{76}$$

With the use of these relations, we have

$$dT^\sigma = \left(\frac{\partial T^\sigma}{\partial s^\sigma}\right)_{P^\alpha,\sigma} ds^\sigma + \left(\frac{\partial T^\sigma}{\partial P^\alpha}\right)_{s^\sigma,\sigma} dP^\alpha + \left(\frac{\partial T^\sigma}{\partial \sigma}\right)_{s^\sigma,P^\alpha} d\sigma$$

$$= \frac{T^\sigma}{c_p^\sigma} ds^\sigma + \left(\frac{\partial T^\sigma}{\partial P^\alpha}\right)_{s^\sigma,\sigma} dP^\alpha - \left(\frac{\partial a}{\partial s^\sigma}\right)_{P^\alpha,\sigma} d\sigma \qquad (77a)$$

where $c_p^\sigma \equiv T^\sigma(\partial s^\sigma/\partial T^\sigma)_{P^\alpha,\sigma}$.

$$d\upsilon^{\alpha\sigma} = \left(\frac{\partial \upsilon^{\alpha\sigma}}{\partial s^\sigma}\right)_{P^\alpha,\sigma} ds^\sigma + \left(\frac{\partial \upsilon^{\alpha\sigma}}{\partial P^\alpha}\right)_{s^\sigma,\sigma} dP^\alpha + \left(\frac{\partial \upsilon^{\alpha\sigma}}{\partial \sigma}\right)_{s^\sigma,P^\alpha} d\sigma$$

$$= \left(\frac{\partial T^\sigma}{\partial P^\alpha}\right)_{s^\sigma,\sigma} ds^\sigma + \left(\frac{\partial \upsilon^{\alpha\sigma}}{\partial P^\alpha}\right)_{s^\sigma,\sigma} dP^\alpha - \left(\frac{\partial a}{\partial P^\alpha}\right)_{s^\sigma,\sigma} d\sigma \qquad (77b)$$

$$da = \left(\frac{\partial a}{\partial s^\sigma}\right)_{P^\alpha,\sigma} ds^\sigma + \left(\frac{\partial a}{\partial P^\alpha}\right)_{s^\sigma,\sigma} dP^\alpha + \left(\frac{\partial a}{\partial \sigma}\right)_{s^\sigma,P^\alpha} d\sigma \qquad (77c)$$

Substitution of Eqs. (77a-c) in Eq. (74b) gives

$$H^\sigma/N_*^\sigma = \frac{T^\beta}{c_p^\sigma}(ds^\sigma)^2 - \left(\frac{\partial \upsilon^{\alpha\sigma}}{\partial P^\alpha}\right)_{s^\sigma,\sigma}^* (dP^\alpha)^2 + 2\left(\frac{\partial a}{\partial P^\alpha}\right)_{s^\sigma,\sigma}^* dP^\alpha d\sigma + \left(\frac{\partial a}{\partial \sigma}\right)_{s^\sigma,P^\alpha}^* (d\sigma)^2 \qquad (78a)$$

from where the matrix of this quadratic form is

$$\mathbf{H}^\sigma_{(s^\sigma,P^\alpha,\sigma)} = N_*^\sigma \begin{pmatrix} \dfrac{T^\beta}{c_p^\sigma} & 0 & 0 \\[2ex] 0 & -\left(\dfrac{\partial \upsilon^{\alpha\sigma}}{\partial P^\alpha}\right)_{s^\sigma,\sigma}^* & \left(\dfrac{\partial a}{\partial P^\alpha}\right)_{s^\sigma,\sigma}^* \\[2ex] 0 & \left(\dfrac{\partial a}{\partial P^\alpha}\right)_{s^\sigma,\sigma}^* & \left(\dfrac{\partial a}{\partial \sigma}\right)_{s^\sigma,P^\alpha}^* \end{pmatrix} \qquad (78b)$$

It is seen from here that entropy fluctuations are independent of the fluctuations of pressure and surface tension, whereas the fluctuations of $\sigma$ and $P^\alpha$ correlate. Calculating the inverse matrix, we can find the correlator $\langle\Delta\sigma\Delta P^\alpha\rangle$, as well as the fluctuations of $\sigma$ and $P^\alpha$, according to the known formula [33] $\langle\Delta\xi_i\Delta\xi_k\rangle = kT^\beta h_{ik}^{-1}$, where $h_{ik}^{-1}$ is the matrix $(\mathbf{H}^\sigma)^{-1}$ element. Thus, one obtains for the surface tension fluctuation

$$\langle(\Delta\sigma)^2\rangle = \frac{kT^\beta}{N_*^\sigma}\left[\left(\frac{\partial a}{\partial \sigma}\right)_{s^\sigma,P^\alpha}^* + \left(\frac{\partial a}{\partial P^\alpha}\right)_{s^\sigma,\sigma}^{*2}\left(\frac{\partial P^\alpha}{\partial \upsilon^{\alpha\sigma}}\right)_{s^\sigma,\sigma}^*\right]^{-1}$$

$$= \frac{kT^\beta}{N_*^\sigma}\left[\left(\frac{\partial a}{\partial \sigma}\right)_{s^\sigma,P^\alpha}^* + \left(\frac{\partial a}{\partial P^\alpha}\right)_{s^\sigma,\sigma}^*\left(\frac{\partial a}{\partial \upsilon^{\alpha\sigma}}\right)_{s^\sigma,\sigma}^*\right]^{-1} \qquad (79)$$

One more set of variables of interest is ($T^\sigma$, $\upsilon^{\alpha\sigma}$, $a$), accordingly, $s^\sigma$, $P^\alpha$, and $\sigma$ are functions of these variables. From equation

$$df^{\sigma} = -s^{\sigma}dT^{\sigma} - P^{\alpha}dv^{\alpha\sigma} - P^{\beta}dv^{\beta\sigma} + \sigma lda \qquad (80)$$

for the Helmholtz free energy per one molecule,

$$\left(\frac{\partial s^{\sigma}}{\partial v^{\alpha\sigma}}\right)_{T^{\sigma},a} = \left(\frac{\partial P^{\alpha}}{\partial T^{\sigma}}\right)_{v^{\alpha\sigma},a}, \quad \left(\frac{\partial s^{\sigma}}{\partial a}\right)_{T^{\sigma},v^{\alpha\sigma}} = -\left(\frac{\partial \sigma}{\partial T^{\sigma}}\right)_{v^{\alpha\sigma},a}, \quad \left(\frac{\partial P^{\alpha}}{\partial a}\right)_{T^{\sigma},v^{\alpha\sigma}} = -\left(\frac{\partial \sigma}{\partial v^{\alpha\sigma}}\right)_{T^{\sigma},a} \qquad (81)$$

Transforming the differentials $ds^{\sigma}$ and $dP^{\alpha}$ with the aid of these relations similarly to Eqs. (77a, b) and then substituting in Eq. (74b), we get the quadratic form with the following matrix:

$$\mathbf{H}^{\sigma}_{(T^{\sigma},v^{\alpha\sigma},a)} = N_*^{\sigma}\begin{pmatrix} \dfrac{c_V^{\sigma}}{T^{\beta}} & 0 & 0 \\[2ex] 0 & -\left(\dfrac{\partial P^{\alpha}}{\partial v^{\alpha\sigma}}\right)^*_{T^{\sigma},a} & \left(\dfrac{\partial \sigma}{\partial v^{\alpha\sigma}}\right)^*_{T^{\sigma},a} \\[2ex] 0 & \left(\dfrac{\partial \sigma}{\partial v^{\alpha\sigma}}\right)^*_{T^{\sigma},a} & \left(\dfrac{\partial \sigma}{\partial a}\right)^*_{T^{\sigma},v^{\alpha\sigma}} \end{pmatrix} \qquad (82)$$

where $c_V^{\sigma} \equiv T^{\sigma}(\partial s^{\sigma}/\partial T^{\sigma})_{v^{\alpha\sigma},a}$. So, temperature fluctuations are statistically independent of fluctuations of $v^{\alpha\sigma}$ and $a$, whereas the fluctuations of the latter two quantities correlate. The fluctuation of $a$ from here is

$$\left\langle (\Delta a)^2 \right\rangle = \frac{kT^{\beta}}{N_*^{\sigma}}\left[\left(\frac{\partial \sigma}{\partial a}\right)^*_{T^{\sigma},v^{\alpha\sigma}} + \left(\frac{\partial \sigma}{\partial v^{\alpha\sigma}}\right)^{*2}_{T^{\sigma},a}\left(\frac{\partial v^{\alpha\sigma}}{\partial P^{\alpha}}\right)^*_{T^{\sigma},a}\right]^{-1} \qquad (83)$$

Eqs. (79) and (83) are simplified (the mentioned correlations vanish) in the approximation of incompressible ($\alpha\sigma$)-layer, $(\partial v^{\alpha\sigma}/\partial P^{\alpha})_{s^{\sigma},\sigma} = 0$, $(\partial v^{\alpha\sigma}/\partial P^{\alpha})_{T^{\sigma},a} = 0$, which can be accepted as a good approximation for a liquid droplet in vapor. It is reasonable to assume that the derivative $(\partial a/\partial v^{\alpha\sigma})_{s^{\sigma},\sigma}$ is finite; hence, $\partial a/\partial P^{\alpha} = (\partial a/\partial v^{\alpha\sigma})(\partial v^{\alpha\sigma}/\partial P^{\alpha}) = 0$ also. Thus, Eqs. (79) and (83) become as follows:

$$\left\langle (\Delta\sigma)^2 \right\rangle = \frac{kT^{\beta}}{N_*^{\sigma}}\left(\frac{\partial \sigma}{\partial a}\right)^*_{s^{\sigma},P^{\alpha}}, \quad \left\langle (\Delta a)^2 \right\rangle = \frac{kT^{\beta}}{N_*^{\sigma}}\left(\frac{\partial a}{\partial \sigma}\right)^*_{T^{\sigma},v^{\alpha\sigma}} \qquad (84)$$

### 4.3. Density fluctuation

Hereafter, the DF is understood as the complex consisting of phases $\alpha$ and $\sigma$ being in equilibrium with each other; the DF corresponding to a noncritical nucleus is considered, i.e. it is not in equilibrium with phase $\beta$. The theory is presented for a droplet in vapor.

Eq. (71) with Eqs. (74a, b) is used for calculating the matrix $\mathbf{H}$ of the work of noncritical droplet formation. Differently from the previous case, the surface tension is not an independent variable now; the condition of DF internal equilibrium leads to Eq. (61b) from which the surface tension is a function of radius and DF temperature, $\sigma = \sigma(R,T)$, hence, the quadratic form $H^\sigma$, Eq. (74b), has only two independent variables. Temperature $T$ is the common variable for $H^\alpha$ and $H^\sigma$. As a second stable variable, we can take $v^\alpha$ for $H^\alpha$ and $v^{\alpha\sigma}$ for $H^\sigma$. Thus, the full set of variables for the work is $(V, v^\alpha, v^{\alpha\sigma}, T)$. With the aid of relations

$$ds^\alpha = \left(\frac{\partial s^\alpha}{\partial v^\alpha}\right)_T dv^\alpha + \left(\frac{\partial s^\alpha}{\partial T}\right)_{v^\alpha} dT = \left(\frac{\partial P^\alpha}{\partial T}\right)_{v^\alpha} dv^\alpha + \frac{c_V^\alpha}{T} dT \tag{85a}$$

$$dP^\alpha = \left(\frac{\partial P^\alpha}{\partial v^\alpha}\right)_T dv^\alpha + \left(\frac{\partial P^\alpha}{\partial T}\right)_{v^\alpha} dT \tag{85b}$$

one obtains

$$H^\alpha = -\left(\frac{\partial P^\alpha}{\partial v^\alpha}\right)_T^* (dv^\alpha)^2 + \frac{c_V^\alpha}{T^\beta}(dT)^2 \tag{86}$$

Similarly,

$$ds^\sigma dT - dv^{\alpha\sigma} dP^\alpha = -\left(\frac{\partial P^\alpha}{\partial v^{\alpha\sigma}}\right)_T^* (dv^{\alpha\sigma})^2 + \frac{c_V^\sigma}{T^\beta}(dT)^2 \tag{87}$$

Considering the remaining term $da\, d\sigma$, we have

$$d\sigma = \left(\frac{\partial \sigma}{\partial R}\right)_T \frac{dR}{dV} dV + \left(\frac{\partial \sigma}{\partial T}\right)_R dT \tag{88a}$$

and

$$N^\sigma da = dA - a\, dN^\sigma = \left(\frac{2}{R} - a\frac{dN^\sigma}{dV}\right) dV \tag{88b}$$

Thus,

$$N_*^\sigma da\, d\sigma = \frac{2\gamma}{3V_*}\left(\frac{\partial \sigma}{\partial R}\right)_T^* (dV)^2 + \frac{2\gamma}{R_*}\left(\frac{\partial \sigma}{\partial T}\right)_R^* dV dT, \quad \gamma \equiv 1 - \frac{R_* a}{2}\left(\frac{dN^\sigma}{dV}\right)_* \tag{89}$$

In the approximation of incompressible both phase $\alpha$ and ($\alpha\sigma$)-layer, the variables $v^\alpha$ and $v^{\alpha\sigma}$ drop out from consideration [13]. Substituting Eqs. (86), (87), and (89) in Eq. (71), we find the desired matrix $\mathbf{H}$ as follows:

$$\mathbf{H} = \begin{pmatrix} -\dfrac{1}{3V_*}\left[P_L^* - 2\gamma\left(\dfrac{\partial \sigma}{\partial R}\right)_T^*\right] & \dfrac{\gamma}{R_*}\left(\dfrac{\partial \sigma}{\partial T}\right)_R^* \\[3mm] \dfrac{\gamma}{R_*}\left(\dfrac{\partial \sigma}{\partial T}\right)_R^* & \dfrac{C_{V*}^\alpha + C_{V*}^\sigma}{T^\beta} \end{pmatrix} \tag{90}$$

where $C_{V*}^\alpha = N_*^\alpha c_V^\alpha$ and $C_{V*}^\sigma = N_*^\sigma c_V^\sigma$ are the heat capacities of phases $\alpha$ and $\sigma$. The derivatives in this equation are given by Eq. (61b) for the DF internal equilibrium.

If phase $\alpha$ is absent, $N_*^\alpha \to 0$, then $C_{V*}^\alpha \to 0$ and the matrix $\mathbf{H}$ is determined by the surface layer properties only; also, $V^\alpha \to 0$ and $V = V^{\alpha\sigma}$ (Fig. 1). Taking into account that $\upsilon^{\alpha\sigma} = \upsilon^{\alpha\sigma}(V)$, we have

$$Ra = RA / N^\sigma = 3V / N^\sigma = 3\upsilon^{\alpha\sigma}, \quad \frac{dN^\sigma}{dV} = \frac{1}{\upsilon^{\alpha\sigma}}\left[1 - \frac{R}{3\upsilon^{\alpha\sigma}(R)}\frac{d\upsilon^{\alpha\sigma}}{dR}\right]$$

The evaluation of the expression in brackets for functions I-III gives practically the same result: it varies from 0.88 to 0.96 for $R$ changing from 0.3 to 1 nm, respectively; $\gamma$ changes sign only at $R$ about 0.1 nm. So, this expression can be put equal to unity in the practical region of nucleation and $\gamma = -1/2$ with sufficient accuracy (as is discussed below, the parameter $a_\infty$ gives much greater uncertainty to $h_{VT}$); it is significant only that $\gamma < 0$. Hence,

$$\mathbf{H} = \begin{pmatrix} -\dfrac{1}{3V_*}\left[P_L^* + \left(\dfrac{\partial\sigma}{\partial R}\right)_T^*\right] & -\dfrac{1}{2R_*}\left(\dfrac{\partial\sigma}{\partial T}\right)_R^* \\[4mm] -\dfrac{1}{2R_*}\left(\dfrac{\partial\sigma}{\partial T}\right)_R^* & \dfrac{C_{V*}^\sigma}{T^\beta} \end{pmatrix} \qquad (91)$$

It should be recalled for comparison that the matrix $\mathbf{H}$ has the form [13]

$$\mathbf{H}^{(CNT)} = \begin{pmatrix} -\dfrac{P_L^*}{3V_*} & 0 \\[4mm] 0 & \dfrac{C_{V*}^\alpha}{T^\beta} \end{pmatrix} \qquad (92)$$

if the surface term $H^\sigma$ is neglected in Eq. (71) (the CNT approximation). It is seen that Eq. (90) takes into account the dependence of surface tension on radius and the distinction of the heat capacities of phases $\alpha$ and $\sigma$. However, the most significant difference from Eq. (92) is the appearance of the off-diagonal elements related to the dependence of surface tension on temperature; the effect of these elements on the kinetics of droplet evolution will be discussed later.

The quadratic form with the matrix $\mathbf{H}$, Eq. (90), can be identically transformed as follows:

$$H(V,T) = h_{VV}(V - V_*)^2 + 2h_{VT}(V - V_*)(T - T^\beta) + h_{TT}(T - T^\beta)^2$$

$$\equiv \frac{\det\mathbf{H}}{h_{TT}}(V - V_*)^2 + h_{TT}(T - T_e)^2 = \frac{\det\mathbf{H}}{h_{TT}}(V - V_*)^2 + h_{TT}(T' - T^\beta)^2 \qquad (93)$$

where

$$T_e(V) = T^\beta - \frac{h_{VT}}{h_{TT}}(V - V_*) \qquad (94a)$$

$$T' = T + \frac{h_{VT}}{h_{TT}}(V - V_*) \qquad (94b)$$

Such transformation with respect to arbitrary stable variables was used in Ref. [15] for normalizing the equilibrium distribution function. $T_e$ is a solution of equation $\partial H / \partial T = 0$, or $\partial W / \partial T = 0$; this fact together with the form of Eq. (93) leads to the conclusion that $T_e$ plays the role of *equilibrium* *temperature* for a noncritical droplet. The distinction of $T_e$ from the vapor temperature $T^\beta$ does not contradict to Eq. (26a), since there is no full equilibrium for a noncritical droplet; this is the "partial equilibrium" – the equilibrium with respect to temperature only. The full equilibrium takes place for the critical nucleus; equation $dW = 0$ leads to the set of equations $\partial W/\partial V = 0$ and $\partial W/\partial T = 0$ resulting in $V = V_*$ and $T_e = T^\beta$, as it must. As noted above, $\partial \sigma/\partial T < 0$; Eq. (91) gives $h_{VT} > 0$. Hence, $T_e > T^\beta$ for subcritical droplets and $T_e < T^\beta$ for postcritical ones due to the dependence of surface tension on temperature; the postcritical droplets with vapor temperature, $T = T^\beta$, are therefore overheated relatively $T_e$. So, $T_e = T^\beta$ for noncritical droplets holds, only if the dependence of surface tension on temperature is neglected, $h_{VT} = 0$.

Eq. (93) shows that temperature $T$ for the droplet of volume $V \neq V_*$ fluctuates around $T_e$, rather than $T^\beta$; i.e. the center of fluctuations is displaced with changing $V$, whereas the rms fluctuation is determined by $h_{TT}$, as in CNT, $\langle (\Delta T)^2 \rangle = kT^\beta/h_{TT} = k(T^\beta)^2/C_{V*}^\sigma$ for Eq. (91). It should be emphasized that the presence of the off-diagonal terms $h_{VT}$ in Eq. (90) does not mean the correlation of volume and temperature; as is known from the theory of fluctuations[33] as well as shown by Eq. (82), volume and temperature do not correlate. We cannot calculate the correlator $\langle \Delta V \Delta T \rangle$ as well as $\langle (\Delta V)^2 \rangle$ here, since the variable $V$ is *unstable* and the integral over it therefore diverges. Calculation of the temperature rms fluctuation is possible due to the transformation given by Eq. (93) which separates these variables.

The elements $h_{VV}$ and $h_{VT}$ in Eq. (91) are evaluated with the help of Eqs. (40-41) and (59), respectively. Evidently, $(\partial \sigma/\partial R)/P_L \leq 0.5$; the equality is achieved at $R = 0$. Replacing the difference $(s^\alpha - s^\sigma)$ in Eq. (59) by its limiting value given by Eq. (60), we have

$$h_{VT} = \frac{1}{2R_*}\left[ a(R_*)\left(1 + \frac{2\widetilde{\delta}(R_*)}{R_*}\right) \right]^{-1} a_\infty \left|\frac{d\sigma_\infty}{dT}\right|_*, \quad a_\infty = \frac{1}{\rho^\sigma \tau} \tag{95}$$

where $\rho^\sigma$ is the mean density of the surface layer and $\tau$ is its thickness. By substituting the liquid bulk density $\rho_0^\alpha = 33.5$ nm$^{-3}$ for $\rho^\sigma$ and assuming $\tau = 1$ nm, one obtains $a_\infty = 0.03$ nm$^2$. However, $\rho^\sigma < \rho_0^\alpha$; on the other hand, $\tau$ may be greater than 1 nm (three intermolecular distances $d$). The estimates for CCl$_4$ of Ref. [24] show that for the density, $(\rho_0^\alpha - \rho)/\rho_0^\alpha < 0.1\%$ at $\tau = 4d$, whereas for the tangential pressure, $(P - P_t)/P = 0.1\%$ is achieved only at $\tau = 86d$. So, the above value of $a_\infty$ can be strongly overestimated.

As is seen from Fig. 4b, $\upsilon^\sigma$ increases with decreasing $R$; hence, the configurational part of the entropy and $s^\sigma$ increase also. Thus, taking the planar limit for $(s^\alpha - s^\sigma)$, we underestimate $h_{VT}$. Employing $a_\infty$ as an adjustable parameter, this effect can be taken into account also.

As shown earlier, inclusion of the droplet temperature into consideration allows us to study nonisothermal effects in nucleation [13, 14, 20, 50-56] and calculate the mean overheat of droplets due to the release of condensation heat [13, 14]. The present theory gives the possibility to modify the description of nonisothermal effects with account for surface effects. As is discussed in Ref. [13], temperature is inherent in the macroscopic approach, though the latter can be reformulated in terms of energy also. Temperature is also used in Ref. [52], however, a great suppression of the nucleation rate was obtained. This error was shown in Ref. [53] to arise from the use of the inappropriate evaporation rate at small cluster size; it is corrected in this approach by using detailed balance in equilibrium.

In Ref. [54], the cluster is described by the number $N$ of monomers and energy $E$. The present approach, in particular, Eqs. (93) and (94a, b) can be represented in this variables also [13]. If phase $\alpha$ is absent and the curvature dependence of $\upsilon^{\alpha\alpha}$ is neglected, then $V = V^{\alpha\alpha} = \upsilon^{\alpha\alpha} N^\sigma$, $V - V_* = \upsilon^{\alpha\alpha}(N^\sigma - N_*^\sigma)$, $dE = C_{V*}^\sigma dT$, and $T - T^\beta = (E - E_*)/C_{V*}^\sigma$. Eqs. (93) and (94a, b) are transformed as follows:

$$H(N^\sigma, E) = h_{NN}(N^\sigma - N_*^\sigma)^2 + 2h_{NE}(N^\sigma - N_*^\sigma)(E - E_*) + h_{EE}(E - E_*)^2$$

$$\equiv \frac{\det \mathbf{H}_{(N,E)}}{h_{EE}}(N^\sigma - N_*^\sigma)^2 + h_{EE}(E - E_e)^2 = \frac{\det \mathbf{H}_{(N,E)}}{h_{EE}}(N^\sigma - N_*^\sigma)^2 + h_{EE}(E' - E_*)^2 \qquad (96)$$

where

$$E_e(N^\sigma) = E_* - \frac{h_{NE}}{h_{EE}}(N^\sigma - N_*^\sigma), \qquad E' = E + \frac{h_{NE}}{h_{EE}}(N^\sigma - N_*^\sigma) \qquad (97)$$

and $h_{NE} = (\upsilon^{\alpha\alpha})^2 h_{VV}$, $h_{NE} = \upsilon^{\alpha\alpha} h_{VT}/C_{V*}^\sigma$, $h_{EE} = h_{TT}/(C_{V*}^\sigma)^2 = 1/T^\beta C_{V*}^\sigma$

The quantity $E_e(N^\sigma)$ has the meaning of the equilibrium energy of a cluster consisting of $N^\sigma$ molecules; it differs from the saddle point value $E_*$ due to the temperature dependence of surface tension again. As is seen from Eq. (96), energy fluctuates around $E_e$ with

$\langle (\Delta E)^2 \rangle = kT^\beta / h_{EE} = kC_{V*}^\sigma (T^\beta)^2$. Thus, $E_e$ corresponds to the quantity $\overline{E}(N^\sigma)$ in Eq. (16) of Ref. [54]. According to energy balance Eq. (108), the energy change of an incompressible critical droplet is $dE = C_{V*}^\sigma dT$; i.e. the dependence of energy on $N^\sigma$ is attributed to the heat capacity $C_V^\sigma(N^\sigma)$ and the latter is taken at the saddle point. As a result, the matrix $\mathbf{H}_{(N,E)}$ in the CNT approximation is diagonal and the theory in the $(N, E)$-representation differs from that in the $(V, T)$-variables only by units [13]. In Ref. [54], the quantity $d\overline{E}(N^\sigma)/dN^\sigma$ differs from $(-h_{NE}/h_{EE})$ given by Eq. (97), apparently, as a result of considering the direct dependence of the cluster energy on $N^\sigma$, $E = E(N^\sigma, T)$. The consequence of this assumption in the context of the present approach is analyzed in Appendix I.

# 5. Application to the kinetics of droplet nucleation

The approach of macroscopic kinetics in the multivariable theory of nucleation [11-16] is based on the macroscopic equations of motion of a nucleus near the saddle point

$$\dot{\xi}_i = -\sum_{i=1}^{k} z_{ij}(\xi_j - \xi_j^*), \quad \mathbf{Z} = \mathbf{DH}/kT^\beta, \; i = 1,...,k \tag{98}$$

and Onsager's reciprocal relations for the matrix $\mathbf{D}$ of diffusivities in the Fokker-Planck equation. A phenomenological *macroscopic* equation of nucleus growth is linearized near the saddle point and thus an equation for the unstable variable $\xi_1$ is obtained. Equations for stable variables are written from physical considerations and with the use of the matrix $\mathbf{D}$ symmetry properties. The steady state nucleation rate [15]

$$I = C_0 \sqrt{\frac{kT^\beta}{2\pi} \left| h_{11}^{-1} \right|} \left| \kappa_1 \right| e^{-\frac{W_*}{kT^\beta}} \tag{99}$$

involves the negative eigenvalue $\kappa_1$ of the matrix $\mathbf{Z}$; $h_{11}^{-1}$ is the element $(1,1)$ of the matrix $\mathbf{H}^{-1}$; $C_0$ is the normalizing factor of the one-dimensional equilibrium distribution function of nuclei.

It was shown earlier both for bubble and droplet nucleation [11-14] that this kinetic approach is fully consistent with the multivariable thermodynamics of a nucleus, when the surface term $H^\sigma$ is not taken into account. One of the manifestations of this self-consistency of the theory is Eq. (98) for temperature yields the energy balance equation, or the first law of thermodynamics for a nucleus. As is seen from the foregoing, the surface term $H^\sigma$ essentially changes the form of the matrix $\mathbf{H}$. So, it is of interest to clarify the effect of $H^\sigma$ on the kinetics of nucleation and check the self-consistency of the theory in this case.

## 5.1. CNT approximation

As shown above, a single component droplet is described by the set ($V$, $T$). Eq. (98) has the following explicit form:

$$\begin{cases} \dot{V} = -z_{VV}(V - V_*) - z_{VT}(T - T^\beta) \\ \dot{T} = -z_{TV}(V - V_*) - z_{TT}(T - T^\beta) \end{cases} \tag{100}$$

The equation for $\dot{T}$ can be also rewritten in the form

$$\dot{T} = b_T \dot{V} - \lambda_{TT}(T - T^\beta) \tag{101}$$

from where $z_{TV} = b_T z_{VV}$ and $z_{TT} = \lambda_{TT} + b_T z_{VT}$. The symmetry condition $d_{VT} = d_{TV}$ for the matrix $\mathbf{D}/kT^\beta = \mathbf{ZH}^{-1}$ determines $b_T$; in the case of Eq. (92) for the matrix $\mathbf{H}$,

$$b_T = \frac{z_{VT}}{z_{VV}} \frac{h_{VV}}{h_{TT}} \tag{102}$$

The macroscopic growth equation is obtained from the mass transfer equation $\dot{N}^\alpha = j_+ - j_-$, where $j_+ = \pi R^2 \beta_m u^\beta P^\beta / kT^\beta$ and $j_-$ are the fluxes of condensing and evaporating molecules, respectively; $\beta_m$ is the condensation coefficient. The flux $j_-$ can be found from the detailed balancing: we have $j_- = j_+ = \pi R^2 \beta_m u(T) P_e(R,T)/kT$ for the vapor which is in equilibrium with the droplet of radius $R$ and temperature $T$, i.e. for the vapor with temperature $T$ and pressure $P_e(R,T)$; $u(T) = \sqrt{8kT/\pi m}$ is the mean thermal velocity of vapor molecules of mass $m$ and $u^\beta = u(T^\beta)$. This is the detailed balance in the "macroscopic form" which is applied to a single droplet. It reflects the fact that evaporation is determined by the intrinsic droplet properties ($T$, $R$) and does not depend on vapor properties, as in the microscopic theory [53, 54], where the detailed balance in its usual form is applied to the equilibrium distribution of clusters.

In the CNT approximation, $V = V^\alpha = v^\alpha N^\alpha$ and $\dot{N}^\alpha = \dot{V}/v^\alpha$. Thus we get

$$\dot{V} = \xi' \left[ P^\beta - \sqrt{\frac{T^\beta}{T}} P_e(R,T) \right], \quad \xi' \equiv \frac{\pi R^2 \beta_m u^\beta v^\alpha}{kT^\beta} \tag{103}$$

where $P_e(R,T)$ is the equilibrium pressure for the droplet of radius $R$ and temperature $T$ which is given by Eq. (35),

$$P_e(R,T) = C \, \mathrm{e}^{\frac{-q_{(\alpha\beta)} + 2v^\alpha \sigma / R}{kT}} \tag{104}$$

whereas $P^\beta$ is the current vapor pressure here. Linearization of Eq. (103) near the saddle point in the case of constant surface tension and with account for

$$\frac{dP_L}{dV} = -\frac{P_L}{3V} \tag{105}$$

gives

$$z_{VV} = \xi h_{VV} \;, \quad z_{VT} = \xi \frac{k}{v^\alpha} \tilde{q} \;, \quad \tilde{q} \equiv \frac{q_{(\alpha\beta)} - kT^\beta / 2 - v^\alpha P_L^*}{kT^\beta} \;, \quad \xi \equiv \xi' v^\alpha P^\beta / kT^\beta \tag{106}$$

From here and Eq. (102),

$$b_T = \frac{kT^\beta \tilde{q}}{c_{V*}^\alpha V_*} \tag{107}$$

and Eq. (101) becomes the energy balance equation mentioned above,

$$C_{V*}^\alpha dT = dE^\alpha = \left[ q_{(\alpha\beta)} - \frac{kT^\beta}{2} - v^\alpha P_L^* \right] dN^\alpha - C_{V*}^\alpha \lambda_{TT}(T - T^\beta) dt \tag{108}$$

where the last term ($t$ is the time) describes the heat exchange between the droplet and vapor according to Newton's law. Hence,

$$\lambda_{TT} = \frac{4\pi R_*^2}{C_{V*}^\alpha} \alpha, \quad \alpha = \beta_\varepsilon (1 - \beta_m) \rho^\beta u_1^\beta (c_V^\beta + k/2) \tag{109}$$

where $\alpha$ is the heat transfer coefficient [50]; $\beta_\varepsilon$ is the thermal accommodation coefficient of vapor molecules, $u_1^\beta = u^\beta / 4$, and $c_V^\beta$ is the heat capacity of vapor. The above equation for $\alpha$ is obtained in the kinetic theory of gases [57] as the difference between the energy $(c_V^\beta + k/2)T^\beta$ carried by a vapor molecule striking the droplet and the energy (at full thermal accommodation) $(c_V^\beta + k/2)T$ carried by this molecule, when it is reflected back to the vapor; the factor $(1 - \beta_m)$ is the fraction of reflected molecules. It should be emphasized that these elementary acts of collision and reflection are not identical to condensation and evaporation; a condensing (evaporating) molecule gives (takes away) the energy equal to the vaporization heat with the correction terms, as is described by the first addend in RHS of Eq. (108).

## 5.2. Surface effects on kinetics

The case of Eq. (91) for the matrix **H** is considered here, i.e. the droplet entirely consists of the surface phase and has the volume $V^\sigma = (4\pi/3)(R^\beta)^3$; the surface tension depends on radius and temperature.

The above procedure of deriving Eq. (103) also can be applied here with replacements $V \to V^\sigma$ and $N^\alpha \to N^\sigma = V^\sigma / \upsilon^\sigma$. In order to go then from $V^\sigma$ to $V$, Eq. (66) is used:

$$\dot{V}^\sigma = \frac{dV^\sigma}{dV} \dot{V}, \quad \frac{dV^\sigma}{dV} = 1 + \eta \frac{\rho^\alpha}{\rho^{\beta\sigma}} + \frac{\rho^\alpha}{\rho^{\beta\sigma}} \left[ \left( 1 + \frac{\delta(R)}{R} \right)^2 \left( 1 + \frac{d\delta(R)}{dR} \right) - 1 \right] \tag{110}$$

Thus,

$$\dot{V} = \xi_\sigma' \left[ P^\beta - \sqrt{\frac{T^\beta}{T}} P_e^\sigma(R,T) \right], \quad \xi_\sigma' \equiv \frac{\pi (R^\beta)^2 \beta_m u^\beta \upsilon^\sigma}{kT^\beta c}, \quad c \equiv \frac{dV^\sigma}{dV} \tag{111}$$

$c = 1.27$ for $R_* = 0.7$ nm and function I. Eq. (51) is relevant here instead of CNT Kelvin's Eq. (104),

$$P_e^\sigma(R,T) = C^\sigma e^{\frac{-q_{(\sigma\beta)} + \sigma_\infty(T)\Psi(R)}{kT}} \tag{112}$$

Linearization of Eq. (111) near the saddle point gives the coefficients $z_{VV}$ and $z_{VT}$, as before. Calculating the derivative $(d\dot{V}/dV)_*$ with the use of Eq. (112), we have

$$\frac{d\Psi(R)}{dV} = -\frac{v^{\alpha\sigma}}{3V}\left[P_L + \left(\frac{\partial\sigma}{\partial R}\right)_T\right]$$

for the considered case of $V = V^{\alpha\sigma}$. Hence,

$$z_{VV} = \xi^\sigma h_{VV} \ , \qquad \xi^\sigma \equiv \xi'_\sigma v^{\alpha\sigma} P^\beta / kT^\beta \tag{113}$$

i.e. the proportionality between $z_{VV}$ and $h_{VV}$ which is inherent in the CNT approximation, Eq. (106), takes place here also (for $\gamma \approx -1/2$).

Further, calculating the derivative $(d\dot{V}/dT)_*$, we get

$$z_{VT} = \xi^\sigma \frac{k}{v^{\alpha\sigma}} \tilde{q}^\sigma ,$$

$$kT^\beta \tilde{q}^\sigma = q_{(\sigma\beta)} - \frac{kT^\beta}{2} - \sigma_\infty(T^\beta)\Psi(R_*) + T^\beta\left(\frac{d\sigma_\infty}{dT}\right)_*\Psi(R_*) \tag{114}$$

The quantity $q_{(\sigma\beta)}$ was assumed constant; Eq. (51) for $P_S^\sigma(T)$ implies constant $q_{(\sigma\beta)}$. If $q_{(\sigma\beta)}$ depends on temperature, it is not difficult to obtain from the Clapeyron-Clausius equation

$$P_S^\sigma(T) = C^\sigma \exp\left[-\frac{q_{(\sigma\beta)}(T)}{kT} + \int_{T^\beta}^T \frac{1}{kT}\frac{dq_{(\sigma\beta)}}{dT}dT\right] \tag{115}$$

However, the use of this equation does not add a new summand to Eq. (114).

Returning to the equations of motion, Eq. (100), we find that the equation for $b_T$ is not so simple now as Eq. (102) due to the presence of the off-diagonal terms $h_{VT}$ in the matrix $\mathbf{H}$ and therefore Eq. (101) loses its clear interpretation given by Eq. (108). In order to retain Eq. (101) with the same form of Eq. (102) and then to find the new form of the energy balance equation, it is necessary to go to new variables $(V, T')$, where $T'$ is given by Eq. (94b). The matrix $\mathbf{C}$ of transition $(V,T) \rightarrow (V,T')$ is

$$\mathbf{C} = \begin{pmatrix} 1 & 0 \\ -h_{VT}/h_{TT} & 1 \end{pmatrix}, \qquad \begin{pmatrix} V - V_* \\ T - T^\beta \end{pmatrix} = \mathbf{C}\begin{pmatrix} V - V_* \\ T' - T^\beta \end{pmatrix} \tag{116}$$

The matrices $\mathbf{H}$ and $\mathbf{Z}$ are transformed according to equations $\mathbf{H}' = \mathbf{C}^T\mathbf{H}\mathbf{C}$, $\mathbf{Z}' = \mathbf{C}^{-1}\mathbf{Z}\mathbf{C}$ and acquire the form

$$\mathbf{H}' = \begin{pmatrix} \dfrac{\det\mathbf{H}}{h_{TT}} & 0 \\ 0 & h_{TT} \end{pmatrix}, \ \mathbf{Z}' = \begin{pmatrix} z_{VV} - \dfrac{h_{VT}}{h_{TT}}z_{VT} & z_{VT} \\ \dfrac{\det\mathbf{H}}{h_{TT}^2}z_{VT} & \dfrac{h_{VT}}{h_{TT}}z_{VT} + z_{TT} \end{pmatrix} \tag{117}$$

where the matrix $\mathbf{D}$ symmetry was used in order to transform the element $z'_{TV}$ to the given form. It is easy to check that this transformation retains the matrix $\mathbf{D}$ symmetry. From this equation, $h'^{-1}_{VV} = h_{TT}/\det\mathbf{H} = h^{-1}_{VV}$; also, the eigenvalue $\kappa_1$ is invariant. Thus the nucleation rate, Eq. (99), is *invariant*, as it must from the physical point of view.

This transformation allows writing Eq. (101) for $T'$:

$$\dot{T}' = b_T^\sigma \dot{V} - \lambda'_{TT}(T' - T^\beta) \tag{118}$$

$$z'_{TV} = b_T^\sigma z'_{VV}, \quad z'_{TT} = \lambda'_{TT} + b_T^\sigma z'_{VT} \tag{119}$$

As the matrix $\mathbf{H}'$ is diagonal now, equation for $b_T^\sigma$ has the form of Eq. (102),

$$b_T^\sigma = \frac{z'_{VT}}{z'_{VV}} \frac{h'_{VV}}{h'_{TT}} \tag{120a}$$

Substituting the elements from Eq. (117), we get

$$b_T^\sigma = \left(1 - \frac{h_{VT}^2}{h_{VV}h_{TT}}\right)\left(1 - \frac{h_{VT}}{h_{VV}}b_T^0\right)^{-1} b_T^0, \quad b_T^0 \equiv \frac{z_{VT}}{z_{VV}} \frac{h_{VV}}{h_{TT}} \tag{120b}$$

where $b_T^0$ has the same form as Eq. (102), however, the elements $z_{VV}$ and $z_{VT}$ are given by Eqs. (113) and (114) now, whereas $h_{VV}$ and $h_{TT}$ are given by Eq. (91):

$$b_T^0 = \frac{k\tilde{q}^\sigma}{\upsilon^{\alpha\alpha}h_{TT}} = \frac{kT^\beta \tilde{q}^\sigma}{\upsilon^{\alpha\alpha}C_{V*}^\sigma} \tag{120c}$$

Substituting the obtained $b_T^\sigma$ in Eq. (118) and returning to the temperature $T$, according to Eq. (94b), we have

$$\dot{T} = \left(1 - \frac{h_{VT}}{h_{VV}}b_T^0\right)^{-1}\left[b_T^0 - \frac{h_{VT}}{h_{TT}}\right]\dot{V} - \lambda'_{TT}(T - T_e) \tag{121}$$

It should be noted that

$$0 < \left(1 - \frac{h_{VT}}{h_{VV}}b_T^0\right)^{-1} = \frac{z_{VV}}{z'_{VV}} < 1 \tag{122}$$

due to $h_{VT} > 0$ and $(-h_{VT}/h_{TT})\dot{V} = \dot{T}_e$. After substitution of $b_T^0$, Eq. (120c), and with account for $\dot{V} = \dot{V}^{\alpha\alpha} = \upsilon^{\alpha\alpha}\dot{N}^\sigma$, Eq. (121) acquires the following final form

$$C_{V*}^\sigma \dot{T} = \frac{z_{VV}}{z'_{VV}}\left[kT^\beta \tilde{q}^\sigma \dot{N}^\sigma + C_{V*}^\sigma \dot{T}_e\right] - C_{V*}^\sigma \lambda'_{TT}(T - T_e) \tag{123}$$

which is the energy balance equation in the considered case. The last term describes the relaxation of droplet temperature $T$ to its equilibrium value, which confirms the above definition of $T_e$ as an equilibrium temperature for the droplet; when the droplet does not grow, $\dot{N}^\sigma = 0$ and $\dot{T}_e = 0$, Eq. (123) gives an equation for this relaxation:

$$\dot{T} = -\lambda'_{TT}(T - T_e) \tag{124}$$

Eq. (123) differs from Eq. (108) for the same reason as temperature $T_e$ differs from $T^\beta$. If the temperature dependence of surface tension is neglected, then $h_{VT} = 0$, $T_e = T^\beta$, $\dot{T}_e = 0$, $z_{VV}/z'_{VV} = 1$, and Eq. (123) becomes similar to Eq. (108). From these considerations, we can identify $\lambda'_{TT}$ with $\lambda_{TT}$ and express it via the heat transfer coefficient, as in the CNT approximation.

Eq. (112) was derived from the condition of equilibrium of phases $\sigma$ and $\beta$, accordingly, it involves the dependences $\sigma(R)_T$ and $\sigma_\infty(T)$ related to the condition of full equilibrium in the system. At the same time, Eq. (91) was derived from the condition of the DF internal equilibrium and involves the corresponding derivatives of the surface tension. As noted above, both these types of equilibrium yield the same equations for these derivatives in the case of a droplet in the vapor. This fact was used, in particular, in deriving Eq. (113); otherwise, it would be necessary to distinguish the dependences $\sigma(R,T)$ obtained under different conditions and the above proportionality between $z_{VV}$ and $h_{VV}$ would not hold.

In order to estimate the nonisothermal and curvature effects, the explicit form of the matrix $\mathbf{Z}$ is needed. The first row elements are given by Eqs. (113) and (114). Comparing the element $z'_{TT}$ from Eqs. (119) and (117), we get

$$z_{TT} = \lambda_{TT} + \left( b_T^\sigma - \frac{h_{VT}}{h_{TT}} \right) z_{VT} \tag{125}$$

The element $z_{TV}$ is obtained from the symmetry condition $d_{VT} = d_{TV}$ of the matrix $\mathbf{D}$ as follows:

$$z_{TV} = \frac{z_{VT} h_{VV} + (z_{TT} - z_{VV}) h_{VT}}{h_{TT}} \tag{126}$$

Eqs. (125) and (126) complete the matrix $\mathbf{Z}$ determination. The negative eigenvalue $\kappa_1$ of this matrix determines the nucleation rate, Eq. (99),

$$\kappa_1 = \frac{1}{2} \left\{ Sp\mathbf{Z} - \sqrt{(Sp\mathbf{Z})^2 - 4 \det \mathbf{Z}} \right\} \tag{127}$$

In the isothermal limit, $\lambda_{TT} \to \infty$,

$$\kappa_1^{iso} = \frac{\det \mathbf{Z}}{Sp\mathbf{Z}} = z_{VV} - \frac{h_{VT}}{h_{TT}} z_{VT} = z'_{VV} \tag{128}$$

Hence, the ratio of the actual nucleation rate $I$ to the isothermal one $I_{iso}$ is

$$\frac{I}{I_{iso}} = \frac{\kappa_1}{z'_{VV}} \tag{129}$$

The ratio of the isothermal nucleation rates of the present theory and CNT is of special interest; from above equations,

$$\frac{I_{iso}}{I_{iso}^{(CNT)}} = \sqrt{\frac{h_{TT} h_{VV}^{(CNT)}}{\det H}} \frac{z'_{VV}}{z_{VV}^{(CNT)}} \exp\left( \frac{W^{(CNT)} - W}{kT^\beta} \right) \tag{130}$$

where $W$ is given by Eq. (29). All the quantities included in this equation are functions of the critical radius.

In Fig. 5a, the dependence of vapor supersaturation on critical radius is presented; the plots for different functions $\delta(R)$ are almost indistinguishable in this scale, therefore only the plot for function

I is shown. It is seen that ideal-gas Eq. (51) and real-gas Eq. (54) give the same dependence, except the vicinity of the point $R = 0$. These equations give a finite (limiting) supersaturation at $R \to 0$, whereas Kelvin Eq. (35) (curve $K$) gives infinite one.

The ratio $I_{iso} / I_{iso}^{(CNT)}$ vs. the supersaturation is plotted in Fig. 5b for different functions $\delta(R)$. This dependence is mainly due to the exponential function in Eq. (130). At the given supersaturation, the present theory and CNT yield different critical radii and different surface tensions, hence, different $W$. Function II for $R_* = 0.72$ nm gives $\ln(P^{\beta}/P_s) = 1.29$ and $\ln(I_{iso}/I_{iso}^{(CNT)}) = 12.9$, i.e. $I_{iso}/I_{iso}^{(CNT)} = 4 \times 10^5$. As is seen, this ratio increases with decreasing supersaturation, or increasing critical radius, at least up to 2 nm. However, such large radii are beyond the region of nucleation, $W(2)/kT^{\beta} = 294$; apparently, the values about 1 nm are ultimate, $W(1)/kT^{\beta} = 71$.

## 5.3. Estimates of nonisothermal effects

For studying the mean overheat of droplets in some region of sizes, the distribution functions are employed [13, 14]. In the dimensionless variables $x = (V - V_*)/V_*$ and $y = (T - T^{\beta})/T^{\beta}$, the matrices $\mathbf{H}$ and $\mathbf{Z}$ change as follows:

$$h_{xx} = h_{VV} V_*^2, \;\; h_{yy} = h_{TT} (T^{\beta})^2, \;\; h_{xy} = h_{VT} V_* T^{\beta} \; ; \;\; z_{xx} = z_{VV}, \;\; z_{yy} = z_{TT}, \;\; z_{xy} = z_{VT} \frac{T^{\beta}}{V_*}, \;\; z_{yx} = z_{TV} \frac{V_*}{T^{\beta}} \quad (131)$$

The multivariable steady state distribution function $f_s(\mathbf{r})$ is [15, 54]:

$$f_s(\mathbf{r}) = \frac{1}{2} f_e(\mathbf{r}) \, erfc\left(\phi(\mathbf{r})\right), \quad \phi(\mathbf{r}) \equiv -\frac{\mathbf{e}\mathbf{H}\mathbf{r}}{\sqrt{2kT^{\beta}|\kappa_b(\mathbf{e})|}} \;\; , \;\; \mathbf{r} = \begin{pmatrix} x \\ y \end{pmatrix} \quad (132)$$

where $\mathbf{e} = (\cos\theta, \sin\theta)$ is the unit vector of the droplets flux direction on the $(x, y)$-plane;

$$\kappa_b(\mathbf{e}) = \frac{h_{xx} + 2h_{xy} \tan\theta + h_{yy} \tan^2\theta}{1 + \tan^2\theta} \quad (133)$$

is the barrier curvature (or the curvature of the normal section of the saddle surface) in this direction. So, the function $\phi(\mathbf{r})$ acquires the form

$$\phi(x, y) = -\frac{(h_{xx} + h_{xy} \tan\theta) x + (h_{xy} + h_{yy} \tan\theta) y}{\sqrt{2kT^{\beta}|h_{xx} + 2h_{xy} \tan\theta + h_{yy} \tan^2\theta|}} \quad (134)$$

Equation for $\tan\theta$ is [15]

$$\tan\theta = \frac{1}{2z_{xy}}\left\{(z_{yy} - z_{xx}) - \sqrt{(z_{yy} - z_{xx})^2 + 4z_{xy} z_{yx}}\right\} \quad (135)$$

The equilibrium distribution function is

$$f_e(x,y) = C_e \, e^{\frac{1}{kT^\beta}\left\{-W_* + \frac{1}{2}\left(h_{xx}|x^2 - 2h_{xy}xy - h_{yy}y^2\right)\right\}} \tag{136}$$

the exact value of the normalizing constant $C_e$ is not required here, as it is canceled in the following relations.

The steady state and equilibrium temperature distributions of droplets in some region $\Omega(\Delta)$ in $x$ are respectively described by the functions [13]

$$f_{s,\Delta}(y) = \frac{\int\limits_{\Omega(\Delta)} f_s(x,y)dx}{\int\limits_{\Omega(\Delta)} dx \int\limits_{-\infty}^{+\infty} dy f_s(x,y)}, \qquad f_{e,\Delta}(y) = \frac{\int\limits_{\Omega(\Delta)} f_e(x,y)dx}{\int\limits_{\Omega(\Delta)} dx \int\limits_{-\infty}^{+\infty} dy f_{eq}(x,y)} \tag{137}$$

where the parameter $\Delta$ characterizes the region width. The value $\Delta = 0.3$ is employed here and three regions are considered: subcritical, $\Omega = [-\Delta, 0]$, near-critical, $\Omega = [-\Delta, \Delta]$, and postcritical, $\Omega = [0, \Delta]$. Hence, the mean steady state and equilibrium overheats of droplets relatively the vapor temperature in the region $\Omega(\Delta)$ are respectively

$$\Delta T_{s,\Delta} = \langle T_{s,\Delta} \rangle - T^\beta = T^\beta \int\limits_{-\infty}^{+\infty} y f_{s,\Delta}(y)dy, \qquad \Delta T_{e,\Delta} = \langle T_{e,\Delta} \rangle - T^\beta = T^\beta \int\limits_{-\infty}^{+\infty} y f_{e,\Delta}(y)dy \tag{138}$$

where $\langle T_{s,\Delta} \rangle$ and $\langle T_{e,\Delta} \rangle$ are the mean steady state and equilibrium temperatures of droplets in the region $\Omega(\Delta)$. The steady state overheat calculated relatively the mean equilibrium temperature is

$$\Delta T_{s,\Delta}^{(e)} = \Delta T_{s,\Delta} - \Delta T_{e,\Delta} = \langle T_{s,\Delta} \rangle - \langle T_{e,\Delta} \rangle \tag{139}$$

Estimates were made for water droplets at $T^\beta = 20\ °C$, $d\sigma_\infty/dT = -0.15$ erg/cm$^2$ K, $R_* = 0.7$ nm, and function I. Fig. 6a shows the ratio of the nucleation rate to the isothermal one, Eq. (129), as a function of the condensation coefficient. Functions $f_{s,\Delta}(y)$ and $f_{e,\Delta}(y)$ for postcritical droplets are plotted in Fig. 6b. It is seen that the steady state function is shifted relatively the equilibrium one to higher temperatures, which means the average steady state overheat relatively the equilibrium one. Fig. 7a shows both the steady state and equilibrium overheat of sub-, near-, and postcritical droplets relatively the vapor temperature; the CNT result at the same conditions is presented for comparison. The same steady state overheat relatively the equilibrium one, Eq. (139), is shown in Fig. 7b. This is the *kinetic part* of the overheat due to the release of the condensation heat.

The mean equilibrium "overheat" for postcritical droplets is negative, as was mentioned above. This quantity is of opposite sign for sub- and postcritical droplets; it is equal to $\pm\ 0.4\ °K$ for $\Delta = 0.3$ and $a_\infty = 0.03$ nm$^2$. The mean equilibrium overheat for a symmetric region (for near-critical droplets) is absent. Each steady state curve approaches its equilibrium value (straight dashed line in Fig. 7a) at $\beta_m \to 0$ (the isothermal limit), similarly as the CNT curves approach $T^\beta$. It was noted above that, probably, the given value of $a_\infty$ is strongly overestimated. Thus, the realistic values of $h_{xy}$ may be much less and therefore the equilibrium overheat may be negligible. However, the results will not coincide with CNT at $h_{xy} \to 0$, since other surface effects still remain, such as different heat capacities and curvature corrections. Fig. 8a shows the steady state overheat of near-critical droplets for different values of heat capacity $c_V^\sigma$. With increasing value of this quantity, the overheat decreases, as it must. It was mentioned in Ref. [13] that the isothermal limit can be formally achieved at $c_V \to \infty$.

In the nonisothermal limit $\beta_m \to 1$, when there is no heat exchange between the droplet and vapor (in the absence of a carrier gas), the mean steady state overheat reaches its maximum value. As was noted earlier [13], the nucleation rate tends to zero in both isothermal and nonisothermal limits, which is shown in Fig. 8b by the dependence $\kappa_1(\beta_m)$, Eq. (127). There are two governing kinetic parameters in the theory: $z_{VV}' \sim \beta_m$ (characterizes the rate of droplet growth) and $\lambda_{TT} \sim (1 - \beta_m)$ (characterizes the rate of heat exchange). When one of these parameters is small, $\kappa_1$ is proportional to it; $\kappa_1 = z_{VV}' \to 0$ at $\beta_m \to 0$ and $\kappa_1 \sim \lambda_{TT} \to 0$ at $\beta_m \to 1$. The conclusion that the nucleation rate goes to zero at $\beta_m \to 1$ distinguishes the present approach from previous treatments of nonisothermal effects [20, 53-56], where $\beta_m = 1$ is assumed; moreover, the assumption of $\beta_m = 1$ is usual in literature on droplet nucleation. This difference is analyzed in Appendix II by comparison of the present approach with the theory of Ref. [20].

## 6. Conclusion

The main shortcoming of CNT – the representation of a new-phase nucleus as a fragment of bulk phase $\alpha$ - is overcame by introducing into consideration the surface phase $\sigma$ with parameters different from the parameters of coexisting bulk phases in the framework of the FTL method. The definition of density fluctuation (DF) consisting of phases $\alpha$ and $\sigma$ being in equilibrium with each other extends the CNT concept of nucleus. The nucleus itself in the given theory is the DF core bounded by the surface of tension; it includes, in addition to bulk phase $\alpha$, the part of the surface layer. This model allows us to consider by the thermodynamic way the cases when phase $\alpha$ is absent and the nucleus entirely consists of phase $\sigma$, up to the limit of $R \to 0$ when only the part of the surface layer remains.

The derived equation for the work of DF formation is basic for the multivariable theory of nucleation. Equality to zero of the first differential of the work yields conventional conditions of equilibrium of the system which determine the parameters of the critical nucleus and allow deriving different relations between the parameters of coexisting phases. In this way, the dependences of surface tension on radius and temperature are considered. These dependences are shown to arise from differences in the specific volumes and entropies of bulk and surface phases. The condition of the DF internal equilibrium at a fixed state of the parent phase is found to give the same dependences of surface tension for a droplet in vapor far from the critical point as the condition of full equilibrium in the system. The derived equation for the equilibrium pressure over a droplet of small size replaces CNT Kelvin's equation and gives a finite value of the vapor supersaturation at zero critical radius, differently from the latter. It is shown that the ratio of the isothermal nucleation rate to that of CNT can achieve several orders of magnitude due to the curvature effect. The analysis of different dependences of the Tolman length on radius, $\delta(R)$, results in conclusions that the curvature effect is determined by the value of $\delta(0)$ which is positive and relates to the limiting (spinodal) supersaturation and the function $\delta(R)$ decreases with increasing $R$. At the same time, this effect is weakly sensitive to the form of the function $\delta(R)$ and its asymptotic value $\delta_\infty$.

The second differential of the work gives the quadratic form represented by a saddle surface in the space of nucleus parameters; the negative term is naturally separated and thereby the nucleus volume is determined as an unstable variable. The remaining summands form two groups corresponding to phases $\alpha$ and $\sigma$. They allow us to calculate (i) the fluctuations of the parameters of phase $\sigma$ (in this way, the fluctuations of surface tension and specific surface area are calculated) and (ii) the fluctuations of nucleus parameters and the work of a noncritical nucleus formation. An explicit expression for this work (the matrix $\mathbf{H}$ of the quadratic form) is obtained for droplets. The surface phase contribution to the work is found to change essentially the matrix $\mathbf{H}$ in comparison to the CNT matrix: (i) the off-diagonal terms related to the dependence of surface tension on temperature appear; (ii) the dependence of surface tension on radius changes the nucleation barrier curvature; (iii) the heat capacity of a nucleus is the sum of contributions from phases $\alpha$ and $\sigma$. As a consequence of the presence of the off-diagonal terms, the defined equilibrium temperature of a noncritical nucleus differs from the vapor temperature.

With the use of thermodynamic results, the macroscopic kinetics of evolution of a single component droplet entirely consisting of the surface phase is considered. The derived energy balance equation for the droplet differs from the corresponding CNT equation due to the dependence of surface tension on temperature. The mean steady state overheat of droplets consists of thermodynamic and kinetic parts; the former is due to the distinction of the droplet equilibrium temperature from the vapor temperature, whereas the latter is due to the release of the condensation heat during the droplet growth.

## Appendix I

Differently from temperature, the cluster energy $E(N,T)$ is an extensive quantity. So, both the variables of cluster description $N$ and $E$ are extensive; hence, the matrix $\mathbf{H}_{(N,E)}$ must include off-diagonal terms [11] (even in the CNT approximation with constant $\sigma$), which is the case of Ref. [54]. In order to obtain its explicit form, we can transform the quadratic form

$$H_{(N,E)} = h_{NN}(dN)^2 + 2h_{NE}dNdE + h_{EE}(dE)^2$$

with account for

$$dN = \frac{1}{\upsilon}dV, \quad dE = \left(\frac{\partial E}{\partial N}\right)_T dN + \left(\frac{\partial E}{\partial T}\right)_N dT = \frac{E'_N}{\upsilon}dV + C_V dT, \quad E'_N \equiv \left(\frac{\partial E}{\partial N}\right)_T \quad (A1)$$

Comparing the result to the quadratic form $H_{(V,T)}$, we get the following equations:

$$h_{NN} + 2E'_N h_{NE} + E'^2_N h_{EE} = \upsilon^2 h_{VV}, \quad h_{NE} + E'_N h_{EE} = \frac{\upsilon}{C_V}h_{VT}, \quad C_V^2 h_{EE} = h_{TT} \quad (A2)$$

from where

$$\mathbf{H}_{(N,E)} = \begin{pmatrix} \upsilon^2 h_{VV} + \dfrac{E'^2_N}{C_V^2}h_{TT} - \dfrac{2\upsilon E'_N}{C_V}h_{VT} & \dfrac{\upsilon}{C_V}h_{VT} - \dfrac{E'_N}{C_V^2}h_{TT} \\[3mm] \dfrac{\upsilon}{C_V}h_{VT} - \dfrac{E'_N}{C_V^2}h_{TT} & \dfrac{1}{C_V^2}h_{TT} \end{pmatrix} \quad (A3)$$

where all the quantities are taken at the saddle point (asterisk is omitted).

From Eq. (A3), similarly to Eq. (97), one obtains

$$E_e(N) = E_* + \left(E'_N - \upsilon C_V \frac{h_{VT}}{h_{TT}}\right)(N-N_*), \quad E' = E + \left(\upsilon C_V \frac{h_{VT}}{h_{TT}} - E'_N\right)(N-N_*) \quad (A4)$$

The equilibrium energy $E_e(N)$ which differs from Eq. (97) by the additional term $E'_N$ can be identified with the quantity $\overline{E}(N)$ in Ref. [54] and the quantity $-H = d\overline{E}(N)/dN$ therein is equal now to $E'_N - \upsilon C_V h_{VT}/h_{TT}$; the term with $h_{VT}$ takes into account the dependence of surface tension on temperature.

In the CNT approximation ($h_{VT} = 0$), Eqs. (A4) become as follows:

$$E_e(N) = E_* + E'_N(N-N_*), \quad E' = E - E'_N(N-N_*) \quad (A5)$$

The first equation is simply the expansion of energy in $N$; Eq. (97) in this case gives $E_e(N) = E_*$. The matrix $\mathbf{H}_{(N,E)}$ still contains off-diagonal terms, so we cannot apply the computational algorithm of subsection 5.1. Passing to new variables $(N,E) \to (N,E')$ with the transition matrix

$$\mathbf{C} = \begin{pmatrix} 1 & 0 \\ E'_N & 1 \end{pmatrix}, \quad \begin{pmatrix} N-N_* \\ E-E_* \end{pmatrix} = \mathbf{C}\begin{pmatrix} N-N_* \\ E'-E_* \end{pmatrix} \quad (A6)$$

similarly to Eqs. (116) and (117), we get

$$\mathbf{H}_{(N,E')} = \begin{pmatrix} \upsilon^2 h_{VV} & 0 \\ 0 & \dfrac{h_{TT}}{C_V^2} h_{TT} \end{pmatrix}, \qquad \mathbf{Z}_{(N,E')} = \begin{pmatrix} z_{NN} + E'_N z_{NE} & z_{NE} \\ \dfrac{\det \mathbf{H}_{(N,E)}}{h_{EE}^2} z_{NE} & -E'_N z_{NE} + z_{EE} \end{pmatrix} \tag{A7}$$

Transforming the equation of motion $\dot{N} = -z_{NN} dN - z_{NE} dE$ with the aid of Eq. (A1), we find

$$z_{NN} = z_{VV} - \frac{E'_N}{\upsilon C_V} z_{VT}, \quad z_{NE} = \frac{1}{\upsilon C_V} z_{VT}, \quad \mathbf{Z}_{(N,E')} = \begin{pmatrix} z_{VV} & \dfrac{1}{\upsilon C_V} z_{VT} \\ \upsilon C_V b_T z_{VV} & \lambda_{TT} + b_T z_{VT} \end{pmatrix} \tag{A8}$$

The obtained matrices $\mathbf{H}_{(N,E')}$ and $\mathbf{Z}_{(N,E')}$ are the same, as in Ref. [13] in the variables $(N,E)$. The diffusion tensor is

$$\mathbf{D}_{(N,E')} = kT^\beta \mathbf{Z}_{(N,E')} \mathbf{H}_{(N,E')}^{-1} = d_{NN} \begin{pmatrix} 1 & b_E \\ b_E & b^2 + b_E^2 \end{pmatrix}, \quad \mathbf{D}_{(V,T)} = \upsilon^2 d_{NN} \begin{pmatrix} 1 & b_E / \upsilon C_V \\ b_E / \upsilon C_V & (b^2 + b_E^2)/\upsilon^2 C_V^2 \end{pmatrix} \tag{A9}$$

$$b_E = \upsilon C_V b_T = kT^\beta \tilde{q} = q - \frac{kT^\beta}{2} - \frac{2\upsilon^\alpha \sigma}{R} \tag{A10}$$

$b^2$ is given in Ref. [13] and discussed below. It is seen that the diffusion tensor calculated in the $(V,T)$-theory coincides with $\mathbf{D}_{(N,E')}$ up to units.

Returning to the variables $(N,E)$, we transform the diffusion tensor according to the law $\mathbf{D}_{(N,E)} = \mathbf{C} \mathbf{D}_{(N,E')} \mathbf{C}^T$. Putting then $E'_N = b_E$, we get

$$\mathbf{D}_{(N,E)} = d_{NN} \begin{pmatrix} 1 & b_E + E'_N \\ b_E + E'_N & b^2 + (b_E + E'_N)^2 \end{pmatrix} = d_{NN} \begin{pmatrix} 1 & 2b_E \\ 2b_E & b^2 + 4b_E^2 \end{pmatrix} \tag{A11}$$

Eq. (A1) for $dE$ is Eq. (10a) for an incompressible droplet ($dV = 0$, where $dV$ relates to the compression/expansion of the droplet, rather than to $dN$), if entropy is considered as a function of $T$ and $V$. The variables $T$ and $N$ are *independent* in Eq. (A1). At the same time, energy balance Eq. (108) connects these variables: $dE = C_V dT = b_E dN$ without the heat-exchange term, from where $E'_N$ can be identified with $b_E$. Hence, the use of Eq. (A1) doubles $dE$ in comparison with its true value and, accordingly, the doubled value of $d_{NE}$ in the diffusion tensor, Eq. (A11), appears.

So, the consideration of the separate dependence of the cluster energy on $N$ is superfluous. The fact that the variables $(N,E')$ give the correct result is explained by Eq. (A5) for $E'$: the term $-E'_N(N - N_*)$ "removes" the dependence on $N$ from energy. The considered issue does not appear in the $(V,T)$-theory.

## Appendix II

The comparison of the present theory with that by Feder et. al. [20] is carried out here in the CNT approximation for simplicity. First of all, it is of interest to find the heat transfer coefficient $\alpha$ in the theory of Ref. [20] from the point of view of the present approach.

Both in Ref. [13] and Ref. [20], Eq. (A9) for the diffusion tensor was obtained with

$$d_{NN} = \begin{cases} A(N)\rho^{\beta}u_1^{\beta}, & \text{Ref.[20]} \\ A(N)\rho^{\beta}u_1^{\beta}\beta_m, & \text{present theory} \end{cases}, \quad b^2 = \begin{cases} \left(c_V^{\beta} + \dfrac{k}{2}\right)k(T^{\beta})^2 \equiv b_{Fed}^2, & \text{Ref.[20]} \\ \beta_{\varepsilon}\dfrac{1-\beta_m}{\beta_m}b_{Fed}^2, & \text{present theory} \end{cases} \quad \text{(A12)}$$

in the absence of a carrier gas; $A(N) = 4\pi R^2$.

Calculating the matrix $\mathbf{Z} = \mathbf{DH}/kT^{\beta}$ and employing equation $z_{EE} = \lambda_{EE} + b_E z_{NE}$ [13], we find

$$\lambda_{EE} = \lambda_{TT} = \frac{d_{NN}h_{TT}}{kT^{\beta}C_{V*}^2}b^2 \quad \text{(A13)}$$

Comparing this equation to Eq. (109) for $\lambda_{TT}$ and substituting $b^2$ from Eq. (A12), we get

$$\alpha = \begin{cases} \rho^{\beta}u_1^{\beta}(c_V^{\beta} + k/2) \equiv \alpha_{Fed}, & \text{Ref.[20]} \\ \beta_{\varepsilon}(1-\beta_m)\alpha_{Fed}, & \text{present theory} \end{cases} \quad \text{(A14)}$$

So, despite the fact that reflected molecules are not considered by Feder et. al., the heat exchange in their theory presents; the heat transfer coefficient corresponds to $\beta_m = 0$, as if all vapor molecules would be reflected, similarly to inert gas molecules. The inconsistency of this theory from the point of view of the present approach is just in this point: the growth kinetics (the coefficient $d_{NN}$) is calculated with $\beta_m = 1$, whereas the heat exchange corresponds to $\beta_m = 0$. This form of $\alpha$ can be interpreted only by the following two assumptions: (i) in addition to changing the droplet energy by $\pm b_E$, condensing/evaporating molecules also participate in the heat exchange between the droplet and vapor, similarly to striking and reflecting molecules; (ii) the numbers of condensing and evaporating molecules are equal. The second condition holds only at the saddle point, where there is no heat exchange; for a growing droplet, it can be accepted only as approximate. The first assumption implies that, in particular, an evaporating molecule leaves the droplet with energy $\varepsilon_{Vap,2} = (c_V^{\beta} + k/2)T$, similarly to a reflected molecule. As is known, only a liquid-phase molecule with sufficiently high kinetic energy $mu_x^2/2 > b_E$ (on the tail of Maxwell's distribution) can evaporate (cf. the evaporation of crystals in Ref. [2]). This energy is expended to overcome the forces of attraction (which can be associated with the heat of vaporization) and to do the accompanied works; the resulting energy of this molecule in vapor phase is $\varepsilon_{Vap,1} = \langle mu_x^2/2 - b_E \rangle + kT$ and can be expressed via the incomplete gamma functions as follows:

$$\frac{\varepsilon_{Vap,1} - kT}{kT} = \frac{1}{kT}\sqrt{\frac{m}{2\pi kT}} \int_{\sqrt{2b_E/m}}^{\infty} \left(\frac{mu_x^2}{2} - b_E\right) e^{-\frac{mu_x^2}{2kT}} du_x = \frac{1}{2\sqrt{\pi}}\left[\Gamma\left(\frac{3}{2},\tilde{q}\right) - \tilde{q}\,\Gamma\left(\frac{1}{2},\tilde{q}\right)\right] \qquad (A15)$$

$\tilde{q} = b_E/kT$. Evaluation for $\tilde{q} = 16.1$ gives $6.5{\times}10^{-9}$ for this expression, i.e. a water molecule leaves the droplet with zero value of the $x$-component of the kinetic energy; only the contribution $kT$ from other two degrees of freedom remains. As a result, $\varepsilon_{Vap,1}/kT = 1$, whereas $\varepsilon_{Vap,2}/kT = 3.7$ for water vapor. While $\varepsilon_{Vap,2}$ varies for different substances, $\varepsilon_{Vap,1}$ is approximately constant. Thus, $\varepsilon_{Vap,1} \neq \varepsilon_{Vap,2}$ and the first assumption is invalid. Energy $\varepsilon_{Vap,2}$ is the energy of just *vapor* molecule, acquired as a result of its interaction with the droplet before the reflection back to the vapor. It is implied that interacting with the droplet, the molecule comes in equilibrium with it (hence, $\varepsilon_{Vap,2} \sim T$), which is the condition of full thermal accommodation [57]; the coefficient $\beta_\varepsilon$ is the degree of this accommodation.

If $\alpha = \alpha_{Fed}$ is used in the present theory, the dependence $\kappa_1(\beta_m)$ has the form shown in Fig. 8b. The mean steady state overheat in Fig. 8a is less, since all molecules participate in the heat exchange; however, the distinction from the present approach is small. The nucleation rate does not approach zero at $\beta_m \rightarrow 1$ (Fig. 6a) and $I/I_{iso} \sim 0.01$ in this limit.

The following microscopic model corresponds to the present macroscopic approach. The frequency of striking molecules is

$$\nu(N) = A(N)\rho^\beta u_1^\beta = \nu_s(N) + \nu_r(N), \quad \nu_s(N) = \beta_m \nu(N), \quad \nu_r(N) = (1 - \beta_m)\nu(N) \qquad (A16)$$

where $\nu_s$ and $\nu_r$ are the frequencies of sticking and reflection, respectively;

$$\nu_s(N_*) \equiv d_{NN}, \quad \nu_r(N_*) = \frac{1 - \beta_m}{\beta_m}d_{NN} \qquad (A17)$$

A sticking molecule causes the transition $E \rightarrow E + \varepsilon$ in cluster energy with probability $\varphi_s(N,\varepsilon)d\varepsilon$, so that $w_s^+(N,\varepsilon) = \nu_s(N)\varphi_s(N,\varepsilon)$ is the frequency of transitions $(N,E) \rightarrow (N+1, E+\varepsilon)$ and $w_s^-(N,E,\varepsilon)$ is the frequency of reverse transitions. Then

$$\dot{N} = \int d\varepsilon_1 w_s^+(N-1,\varepsilon_1) - \int d\varepsilon_2 w_s^-(N,E,\varepsilon_2) \qquad (A18)$$

Applying detailed balance for the equilibrium distribution function $f_e(N,E) \sim \exp(-W(N,E)/kT^\beta)$,

$$f_e(N-1, E-\varepsilon_2)w_s^+(N-1,\varepsilon_2) = f_e(N,E)w_s^-(N,E,\varepsilon_2) \qquad (A19)$$

and expanding the exponential function, we obtain

$$w_s^-(N,E,\varepsilon_2) = w_s^+(N-1,\varepsilon_2)\left[1 + \frac{W(N,E) - W(N-1, E-\varepsilon_2)}{kT^\beta}\right] \qquad (A20)$$

Expansion of the work gives

$$W(N,E) - W(N-1, E-\varepsilon_2) = \frac{\partial W}{\partial N} + \frac{\partial W}{\partial E}\varepsilon_2 \qquad (A21)$$

Calculating integrals in Eq. (A18) and replacing then $w_s^+(N-1,\varepsilon)$ by $w_s^+(N_*,\varepsilon)$, we get

$$\dot{N} = -\frac{d_{NN}}{kT^\beta}\left[\frac{\partial W}{\partial N} + \frac{\partial W}{\partial E}\langle\varepsilon\rangle_s\right], \quad \langle\varepsilon\rangle_s = \int \varepsilon_2 \varphi_s(N_*,\varepsilon_2)d\varepsilon_2 \tag{A22}$$

where $\langle\varepsilon\rangle_s$ is the mean change of cluster energy when one molecule sticks.

Similarly, a non-accommodating molecule causes the transition $E \to E+\varepsilon$ in cluster energy with probability $\varphi_r(N,\varepsilon)d\varepsilon$ and $w_r^+(N,\varepsilon) = \nu_r(N)\varphi_r(N,\varepsilon)$ is the frequency of these transitions. After interaction with cluster, this molecule is reflected to the vapor making the reverse transition in cluster energy with frequency $w_r^-(N,E,\varepsilon)$. For cluster energy change, we have

$$\dot{E} = \int \varepsilon_1 w_s^+(N-1,\varepsilon_1)d\varepsilon_1 - \int \varepsilon_2 w_s^-(N,E,\varepsilon_2)d\varepsilon_2 + \int \varepsilon_1' w_r^+(N,\varepsilon_1')d\varepsilon_1' - \int \varepsilon_2' w_r^-(N,E,\varepsilon_2')d\varepsilon_2' \tag{A23}$$

Detailed balance for non-accommodating molecules is

$$f_e(N,E-\varepsilon_2')w_r^+(N,\varepsilon_2') = f_e(N,E)w_r^-(N,E,\varepsilon_2') \tag{A24}$$

After application of the above procedure, Eq. (A23) becomes as follows:

$$\dot{E} = -\frac{d_{NN}}{kT^\beta}\left[\frac{\partial W}{\partial N}\langle\varepsilon\rangle_s + \frac{\partial W}{\partial E}\langle\varepsilon^2\rangle_s\right] - \frac{d_{NN}}{kT^\beta}\frac{1-\beta_m}{\beta_m}\frac{\partial W}{\partial E}\langle\varepsilon^2\rangle_r$$

$$= -\frac{d_{NN}}{kT^\beta}\left[\frac{\partial W}{\partial N}\langle\varepsilon\rangle_s + \frac{\partial W}{\partial E}\left(\langle\varepsilon^2\rangle_s + \frac{1-\beta_m}{\beta_m}\langle\varepsilon^2\rangle_r\right)\right] \tag{A25}$$

From Eqs. (A22) and (A25), the diffusion tensor is determined as

$$\mathbf{D} = d_{NN}\begin{pmatrix} 1 & \langle\varepsilon\rangle_s \\ \langle\varepsilon\rangle_s & \frac{1-\beta_m}{\beta_m}\langle\varepsilon^2\rangle_r + \langle\varepsilon^2\rangle_s \end{pmatrix} \tag{A26}$$

Further, the question of determining $\langle\varepsilon\rangle_s$, $\langle\varepsilon^2\rangle_s$, and $\langle\varepsilon^2\rangle_r$ arises. The quantity $\langle\varepsilon\rangle_s$ can be determined either in statistical or thermodynamic way; the former implies the knowledge of the function $\varphi_s(N,\varepsilon)$ for a cluster. From thermodynamic point of view, we know that condensation is accompanied by the release of heat $q$ and by doing two works – the work of cluster surface increase and the work due to the difference in molecular volumes in vapor and liquid [13]. All these contributions to the cluster energy change are taken into account by Eq. (A10) which is the *first law of thermodynamics* for this process. Thus, we can put $\langle\varepsilon\rangle_s = (\delta E)_{therm} = b_E$. This result can be formally attributed to the $\delta$-shaped distribution function $\varphi_s(N,\varepsilon) = \delta(\varepsilon - b_E)$; hence, $\langle\varepsilon^2\rangle_s = b_E^2$. From this point of view, the representation of the cluster energy change in Ref. [20] as $(b_E + x)$, where $x$ is the deviation of the vapor molecule energy from its mean, seems to be inappropriate as a "mixing" of thermodynamic and statistical approaches; the addition of $x$ to $b_E$ does not agree with the first law of thermodynamics, since $x$ is neither heat nor work.

As noted above, the coefficient $\alpha$ is obtained in the kinetic theory of gases [57] from calculating the net energy flux between two plates with different temperatures (here the droplet and vapor play the role of these plates); the term $b^2$ is calculated then with the use of $\alpha$ [13]. So, comparing Eq. (A26) to Eqs. (A9) and (A12), we get

$$\left\langle \varepsilon^2 \right\rangle_r = \beta_\varepsilon \left( c_v^\beta + \frac{k}{2} \right) k (T^\beta)^2 \qquad (A27)$$

The stochastic changes of cluster energy due to interactions with non-accommodating molecules are in agreement with general statistical nature of the energy fluctuations of a subsystem as a result of its thermal interaction with thermostat.

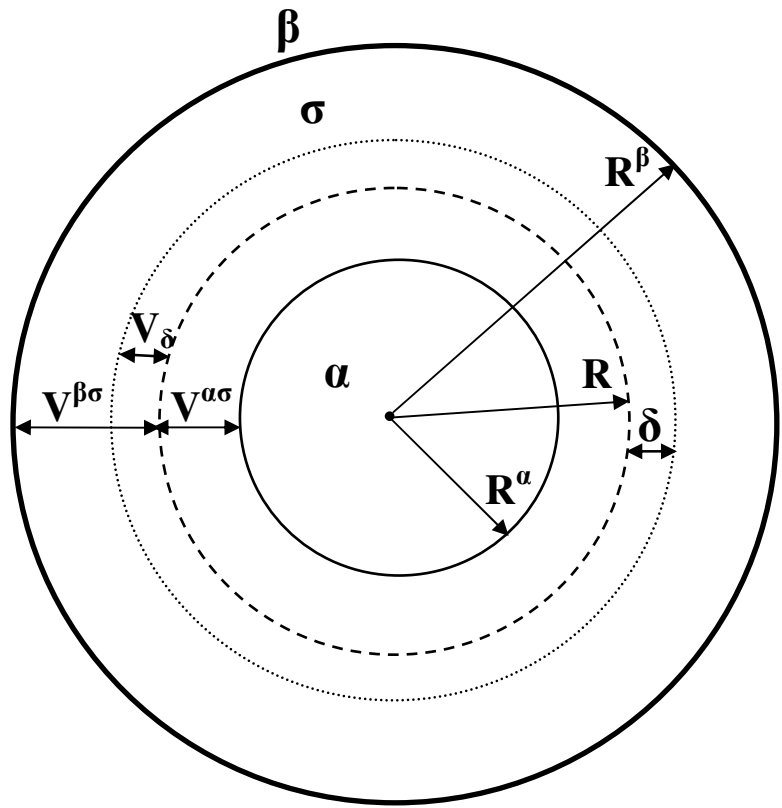

Fig. 1. Phases $\alpha$, $\sigma$, and $\beta$. The density fluctuation is bounded by bold line; the surface of tension and the equimolecular surface are shown by dashed and dotted lines, respectively.

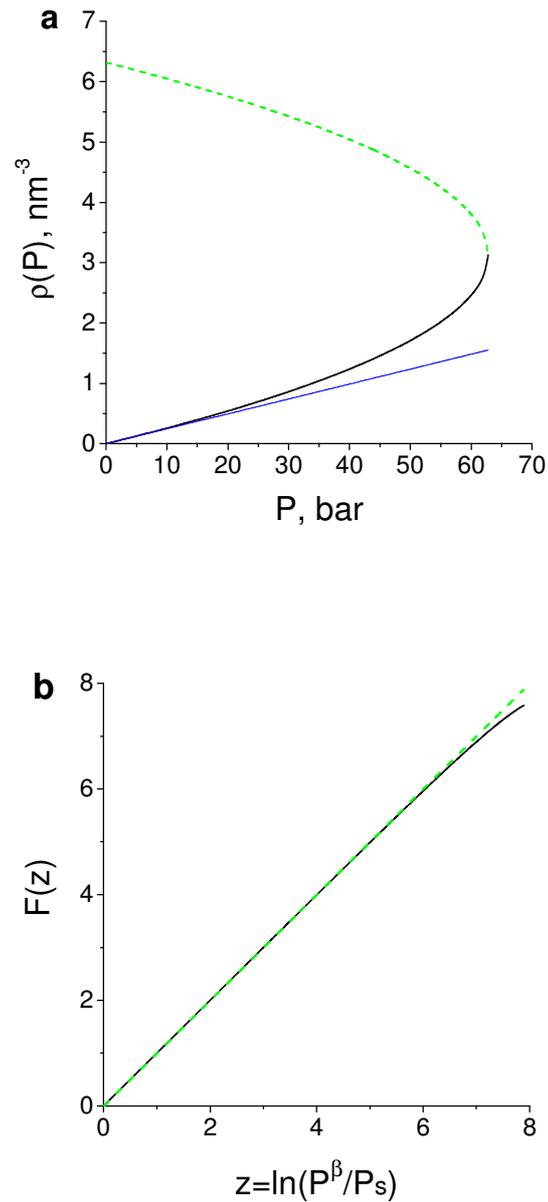

Fig. 2. (a) Dependence of vapor density on its pressure (solid) according to the Van der Waals equation. Dashed line is the root corresponding to the unstable branch of the isotherm; the third root corresponding to the liquid density is outside the given scale. Straight line is the ideal gas dependence. (b) Plot of the function $F(z)$, Eq. (54), (solid) and the corresponding ideal gas dependence $F(z) = z$ (dashed).

**a**

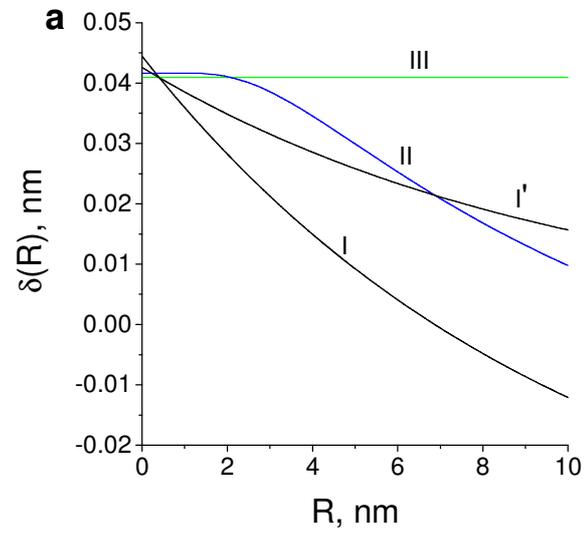

**b**

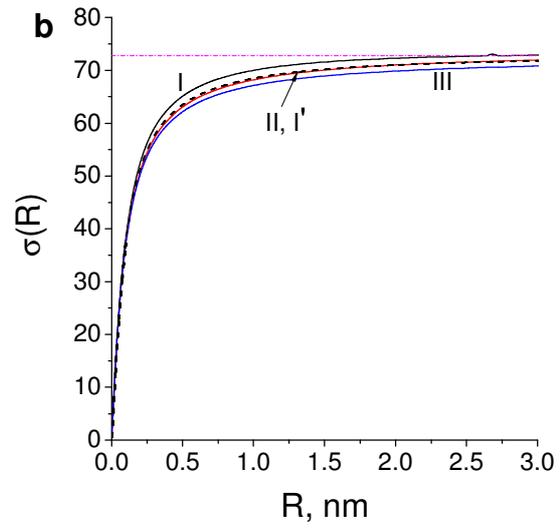

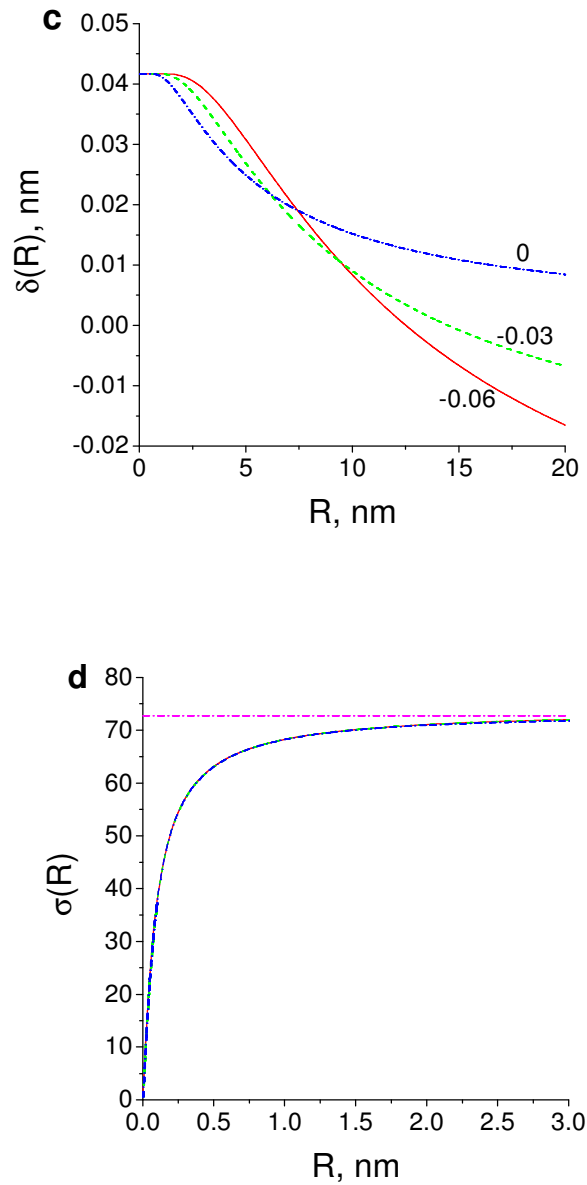

Fig. 3. (a) Functions I, I', II, and III, Eqs. (62a-d); (b) the dependences $\sigma(R)$ corresponding to these functions; dash-dotted line is $\sigma_\infty$ for water. Dashed line relates to function I'. (c) Three variants of function II with the same value of $\delta(0)$ and different values of $\delta_\infty$ shown at the curves; (d) the dependences $\sigma(R)$ corresponding to these functions.

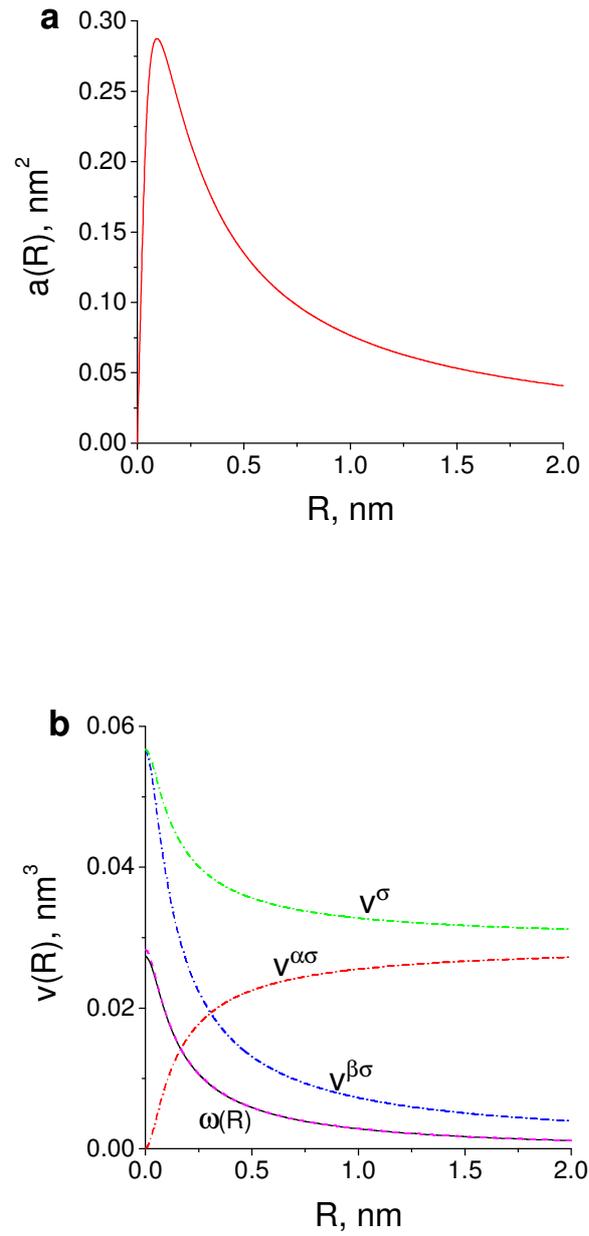

Fig. 4. (a) Dependence $a(R)$. (b) Specific volumes (dash-dotted) and $\omega(R)$, Eqs. (38) (solid) and (49) (dashed). All the plots are for function I.

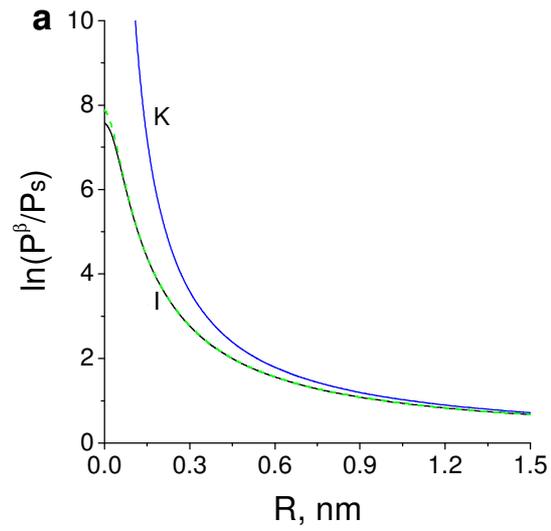

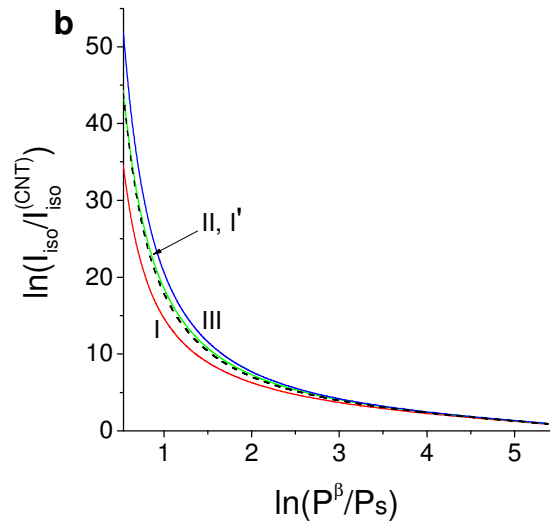

Fig. 5. (a) Vapor supersaturation vs. the droplet critical radius. Solid and dashed curves I relate respectively to Eqs. (51) and (54) and shown for function I; curve K represents Kelvin Eq. (35). (b) The ratio of the isothermal nucleation rates of the present theory and CNT vs. the vapor supersaturation for different functions $\delta(R)$; dashed line relates to function I'.

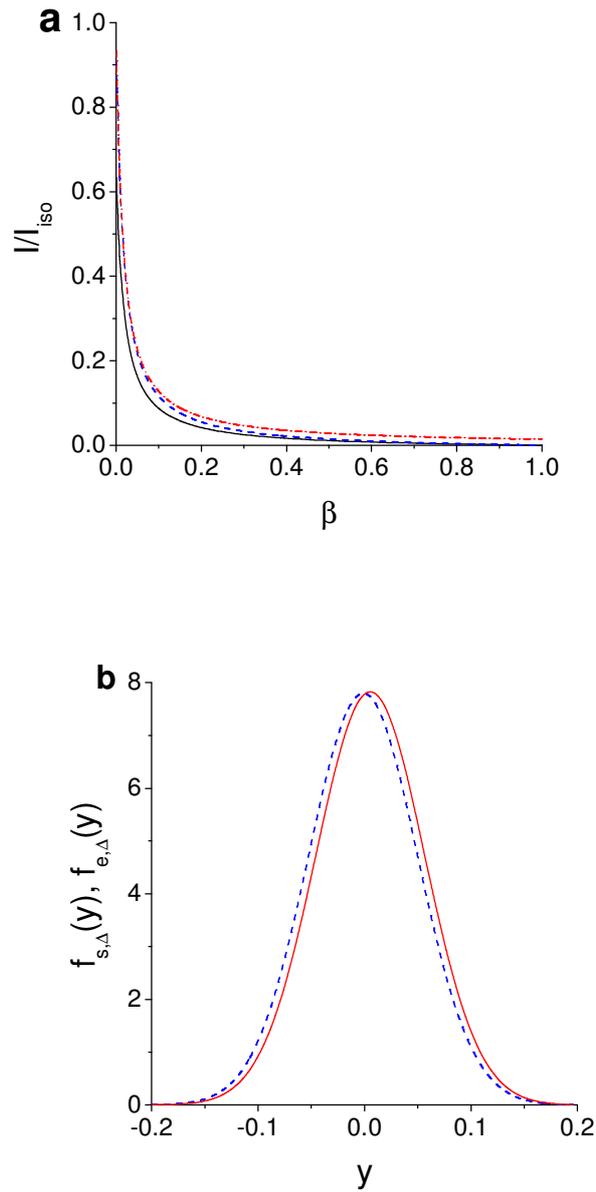

Fig. 6. (a) The ratio of nonisothermal and isothermal nucleation rates as a function of the condensation coefficient in the present theory for $c_V^\sigma = 0.7 c_V^\alpha$ and $a_\infty = 0.03 \ \mathrm{nm}^2$ (solid), in the CNT approximation (dashed), and for the heat exchange coefficient $\alpha = \alpha_{Fed}$, Eq. (A14) - dash-dotted. (b) The steady state (solid) and equilibrium (dashed) temperature distribution of postcritical droplets, $[0, \Delta]$, for the same parameters and $\beta_m = 0.04$.

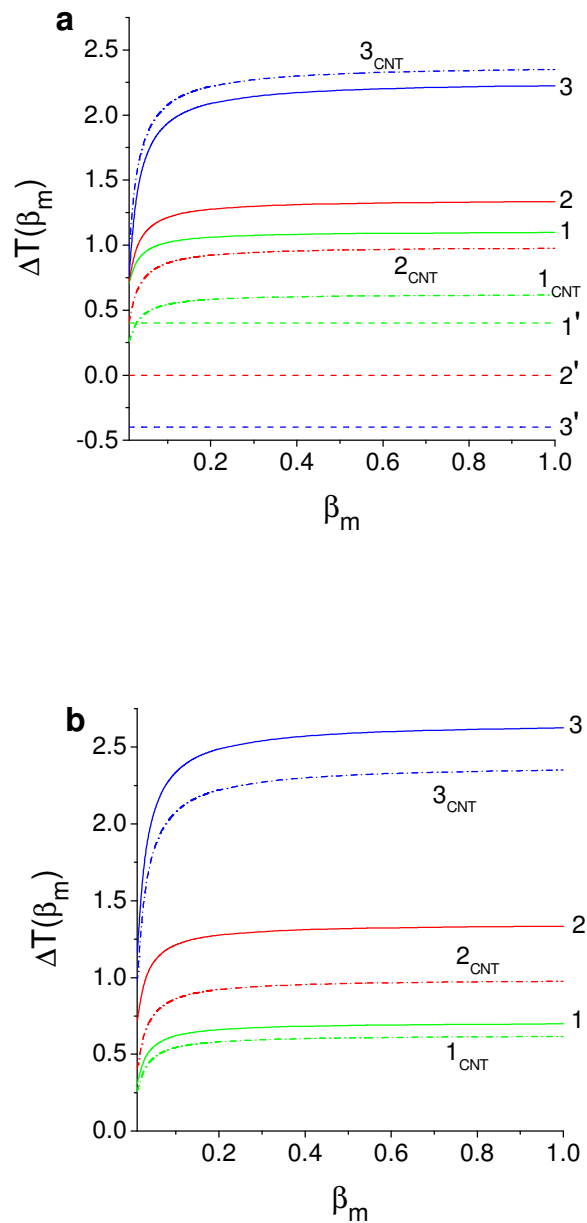

Fig. 7. (a) The mean steady state (solid) and equilibrium (dashed straight lines) overheat of droplets relatively the vapor temperature in the present theory with $c_V^\sigma = 0.7 c_V^\alpha$ and $a_\infty = 0.03$ together with the CNT overheat (dash-dotted). Curves 1, 2, and 3 relate to subcritical, $[-\Delta, 0]$, near-critical, $[-\Delta, \Delta]$ and postcritical, $[0, \Delta]$, droplets, respectively. (b) The same steady state overheat calculated relatively the corresponding equilibrium overheat.

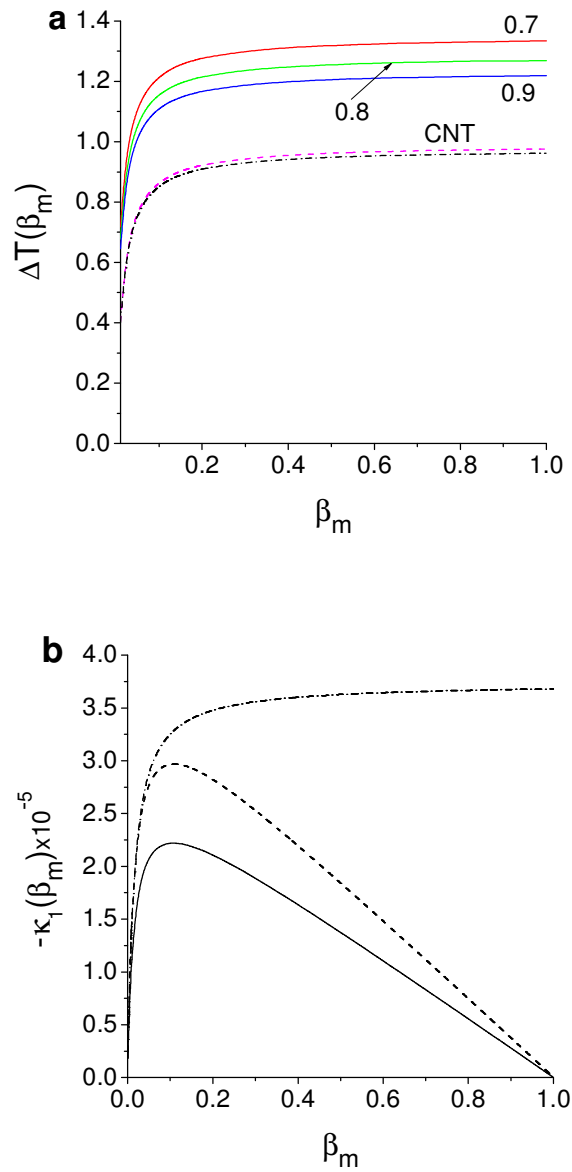

Fig. 8. (a) The mean steady state overheat of near-critical droplets, $[-\Delta, \Delta]$, for $a_\infty = 0.03$ and different values of $c_V^\sigma / c_V^\alpha$ shown at the curves (solid), in CNT (dashed), and for $\alpha = \alpha_{Fed}$, Eq. (A14) - dash-dotted. (b) The dependence $\kappa_1(\beta_m)$ in the present theory (solid) for the parameters of Fig. 6a, in CNT (dashed), and for $\alpha = \alpha_{Fed}$ (dash-dotted).